# 8. Free-space optical links for space communication networks

Alberto Carrasco-Casado[1], Ramon Mata-Calvo[2]

1. Space Communications Laboratory
   NICT – National Institute of Information and Communications Technology, Tokyo (Japan)
2. Institute of Communications and Navigation
   DLR – German Aerospace Center, Oberpfaffenhofen-Wessling (Germany)

**Abstract**

Future spacecraft will require a paradigm shift in the way the information is transmitted due to the continuous increase in the amount of data requiring space links. Current radiofrequency-based communication systems impose a bottleneck in the volume of data that can be transmitted back to Earth due to technological as well as regulatory reasons. Free-space optical communication has finally emerged as a key technology for solving the increasing bandwidth limitations for space communication while reducing the size, weight and power of satellite communication systems, and taking advantage of a license-free spectrum. In the last few years, many missions have demonstrated in orbit the fundamental principles of this technology proving to be ready for operational deployment, and we are now witnessing the emergence of an increasing number of projects oriented to exploit space laser communication (lasercom) in scientific and commercial applications. This chapter describes the basic principles and current trends of this new technology.

**Table of Contents**:

# 8 Free-space optical links for space communication networks

In recent years, wireless communications have witnessed an unprecedented explosion. From cellular networks to satellite links, unguided telecommunications have enabled countless new services, firmly establishing as a basic part of the current information society. In particular, Free-Space Optical Communication (FSOC), despite its recent emergence, provides a number of advantages that allow the materialization of completely-new applications such as quantum communications, as well as the promise to revolutionize traditional applications like satellite communications. FSOC can be applied in a wide variety of scenarios, from crosslinks to up-and-downlinks between satellites, aircraft, ships, and ground standing or mobile terminals. This chapter focuses on communication links where one of the terminals is in space, and it is organized as follows: Section 8.1 gives a quick overview of key concepts of FSOC that help understand how this technology is used to design lasercom links. Section 8.2 describes how the laser-communication signals are affected as they propagate through the atmosphere. Sections 8.3 and 8.4 explain the two applications with more potential, each of them having unique characteristics and advantages. Finally, section 8.5 provides an insight of what optical satellite networks may progress towards in the future.

## 8.1 Principles of free-space optical communication

This section reviews the most fundamental parameters in a space laser communication (lasercom) link. Basically, any space lasercom system will encode some information on a laser beam, collimate and transmit it by means of a telescope, and, after been propagated through free space, it will be collected in other distant telescope and focused on a small spot in the focal plane, where a photodetector will transform the optical signal into an electrical one, which will be decoded to extract the original information. A typical space-lasercom link (see Fig. 8.1) exhibits three different types of impairments, representative of its free-space nature, i.e. geometrical and pointing losses, degradation of the SNR (Signal-to-Noise Ratio) with the background noise, and losses and perturbation of the received signals due to atmospheric effects. This section reviews the first two types, leaving the atmospheric effect to section 8.2 due to its complexity and importance in space lasercom. Other fundamentals concepts specific of free-space links such as modulation, coding, sensitivity, etc. are introduced in this section, as well as a link-budget calculation as the basic design tool in any lasercom system.

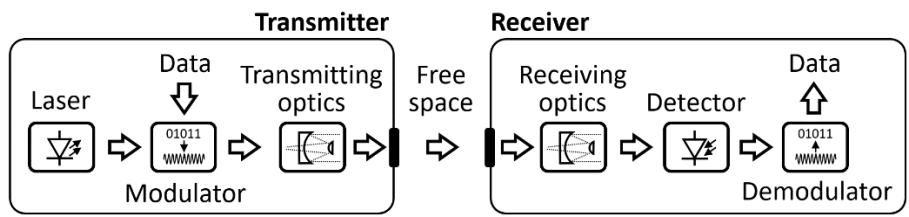

Fig. 8.1. Basic diagram of a generic lasercom system.

### 8.1.1 Brief historical overview

While optical-fiber communications experienced a huge growth in the 1970s, the effects of the atmosphere on the optical signals made the development of FSOC slow down for two decades. In 1992, NASA carried out the first laser transmission to space with the GOPEX project, by emitting 532-nm pulses with a power of MW from the Earth and detecting them with a camera onboard Galileo probe up to 6 million km [1]. In 1994, NICT carried out the first demonstration of a space-to-ground communication downlink by using the Japanese geostationary satellite ETS-VI [2]. After 2000s, there was an accelerated succession of important milestones, being especially noteworthy the following ones: the first inter-satellite link between the ESA's ARTEMIS geostationary satellite and the French LEO satellite SPOT-4 in 2001 [3]; the first link between a High-Altitude Platform (HAP) and ground by DLR in 2005 [4]; the first LEO-to-ground downlink by JAXA and NICT using the OICETS satellite in 2006 [5]; the first link between a satellite (ARTEMIS) and an airplane in 2006 [6]; the first LEO-to-LEO link using Tesat's LCT terminals onboard a US and a German satellite in 2007 [7]; the first aircraft-to-ground link by DLR in 2008 [8]; the first deep-space to ground link using the LADEE probe in orbit around the Moon in 2013 [9]; the first high-speed (over Gbit/s) GEO-to-LEO link between the European satellites Alphasat and Sentinel 1a in 2014 [10]; the first LEO-to-ground lasercom and QKD experiments using a microsatellite (SOTA) by NICT in 2014 [11]; the first link between balloons by Google in 2015 [12]; the first ground experiments in GEO-equivalent scenarios achieving 1.72 Tbit/s by DLR in 2016 [13], and the first quantum-entanglement experiment from space by China using the Micius LEO satellite in 2017 [14].

8.1.2 Key parameters

Diffraction limit

The spot size in the receiver's focal plane and the divergence of the transmitted beam are fundamental parameters in the design of a lasercom system. Ideally, both should be as small as possible: The former gives an idea of how well the receiving optical system can focus the received laser signal, and the later gives an idea of how narrow the transmitting system can transmit a laser beam. Both can be studied by using the concept of diffraction limit, which illustrates one of the best benefits of using optical wavelengths. In optics, the Airy disc has been traditionally used for characterizing the spot size in the receiver's focal plane because it could be easily measured. Derived from the classical description of a plane wavefront illuminating an aperture homogeneously, the Airy disc is defined by the size of the first ring where the light intensity goes to zero. In a more general way, the maximum spatial resolution of a telescope is given by the diffraction limit, which is determined by the wave nature of light and the finite character of the aperture of an optical system. If the aperture's diameter is $D$ and the wavelength is $\lambda$, the angular variation of the intensity of the radiation $I(\theta)/I(0)$ is given by the equation (8.1).

$$\frac{I(\theta)}{I(0)} = \left[ 2 \frac{J_1\left(\frac{\pi D}{\lambda} sin(\theta)\right)}{\frac{\pi D}{\lambda} sin(\theta)} \right]^2 \qquad (8.1)$$

In the equation (8.1), $J_1(x)$ is the Bessel function of the first kind of $x$. Its first minimum corresponds to $x$ = 3.83, or $x$ = 1.61 when $I(\theta)/I(0)$ = -3 dB (see Fig. 8.2). Using the first-minimum criterion and the approximation $sin(\theta) \approx \theta$, the diffraction limit of a telescope can be approximated by the equation (8.2). Sometimes, this limit is calculated assuming other intensity levels to define the first lobe of the Bessel function as shown in the Fig. 8.2. For example, taking the point where the intensity falls to a half, the multiplying factor would be 1.03 in equation (8.2), instead of 1.22 taking the first minimum. Regardless of the convention, in the graphical representation of equation (8.1) shown in the Fig. 8.2, it can be observed that the width of the main lobe of the Bessel function is proportional to the aperture $D$ and inversely proportional to the wavelength $\lambda$. This expression gives an idea of the minimum beam divergence that a perfect telescope, i.e. diffraction limited, can produce.

$$\theta = 1.22\frac{\lambda}{D} \qquad (8.2)$$

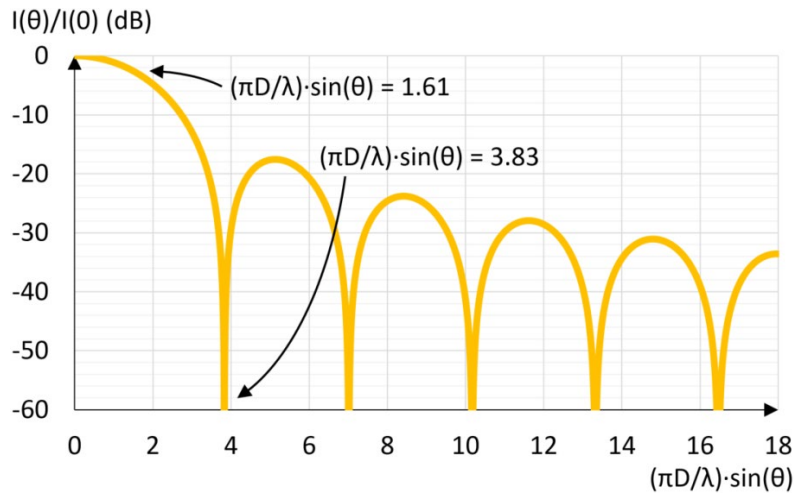

Fig. 8.2. Intensity of radiation as a function of the wavelength $\lambda$, the aperture diameter $D$ and the angular width $\theta$.

The diffraction limit in equation (8.2) represents both the radius of the Airy disk in the focal plane of the receiver's telescope and the minimum beam divergence that a given telescope can ideally produce at a certain wavelength, i.e. in a diffraction-limited system. When the equivalent focal length $f$ of the optical system is considered, the diameter $d$ of the Point Spread Function (PSF) can be calculated by the equation (8.3). The PSF is used to quantify the quality of an optical system and it is defined by the spatial response of the system in the focal plane to a point source in the infinite, which is equivalent to a plane wavefront illuminating the telescope aperture. This expression gives an idea of the minimum spot size that a perfect telescope, i.e. diffraction limited, can produce in the focal plane.

$$d = 2.44\frac{\lambda f}{D} \qquad (8.3)$$

Pointing and tracking

From the transmitter's point of view, equation (8.2) illustrates one of the main advantages of FSOC, i.e. the short wavelengths of light can produce very narrow beams, with minimal divergence, where the energy is well confined. In the case of very long distances, as it is the case in space links, a low divergence becomes a critical factor, allowing a bigger density of power per unit of surface area reaching the receiver. This means that much more power can be delivered to the receiver compared with RF, where the wavelength is much longer. Fig. 8.3 compares an RF and an optical link from Neptune to Earth transmitting with a 40-cm telescope/antenna at a wavelength of 1 μm (IR band) in optical and a frequency of 30 GHz (Ka band) in RF, equivalent to a wavelength of 1 cm. The laser (optical) beam reaching the Earth has a size of around one terrestrial diameter, whereas the RF beam has around 10,000 times the Earth's diameter. This great directivity demands a high pointing accuracy. After the acquisition, when both terminals establish the line of sight to each other, the procedure to keep pointing and tracking is several orders of magnitude more complex than with RF. In RF, the pointing accuracy is in the order of milliradians in the Ka band, whereas a deep-space lasercom link would typically require sub-microradian accuracy.

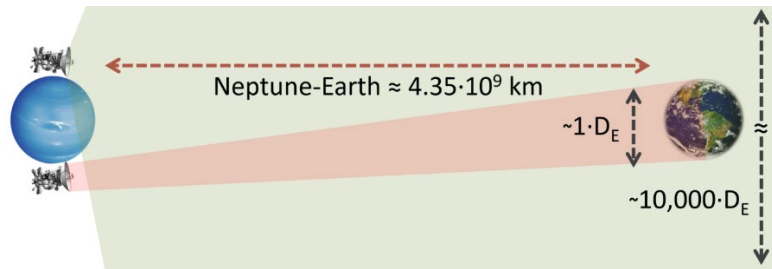

Fig. 8.3. Beam divergence in lasercom and RF from Neptune.

To keep a stable line of sight, it is necessary to use some reference to the other end. This can be achieved by a laser transmitted as a beacon from the ground terminal if the satellite is close to the earth, or celestial references if it is in deep space. The Fig. 8.4 shows the main elements in a typical near-to-Earth link. To initiate the acquisition, the beacon is transmitted with a divergence as wide as the uncertainty zone where the satellite is predicted to be according to its orbital elements. Afterwards, the space system searches for the beacon, looking at the predicted direction of the optical ground station (OGS) and transmitting its downlink towards the beacon at a different wavelength or polarization, once it has been found. Lastly, the OGS can transmit a beam much narrower than the beacon by using the downlink reference in a close loop. Alternatively, scanning algorithms can be implemented, where both terminals scan angularly the counter partner.

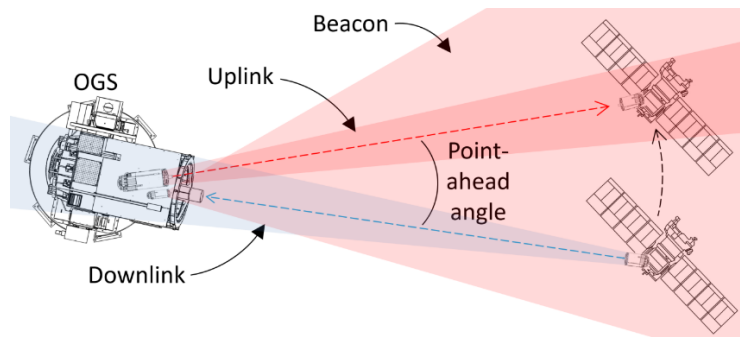

Fig. 8.4. Basic diagram of pointing and tracking.

In space lasercom, the so-called point-ahead angle needs to be considered due to the finite speed of light. Since it takes some time for the uplink beam to reach the moving satellite, both downlink and uplink directions are angularly separated by the point-ahead angle, when this is comparable or larger than the beam width. This angle ɸ can be calculated with the equation (8.4), where $v_t$ is the tangential velocity of the satellite and $c$ is the speed of light.

$$\phi = \frac{2v_t}{c} \qquad (8.4)$$

For a circular orbit, the worst-case point-ahead angle can be obtained by considering the point where the tangential velocity is the fastest, i.e. the zenith. In this case, the tangential velocity is the same as the orbital velocity $v_o$, which can be calculated with equation (8.5), where $G$ is the gravitational constant, $M$ is the mass of the Earth, and $L$ is the distance from the center of the Earth to the satellite. As a reference, the point-ahead angle ɸ is approximately 51 µrad for 500-km LEO and 18 µrad for GEO.

$$v_o = \sqrt{\frac{GM}{L}} \qquad (8.5)$$

Sky radiance

The atmospheric scattering is an effect of dispersion of light originated by particles suspended in the air. Depending on the size of these particles, scattering can be classified in two categories: Rayleigh, when the particles are much smaller than the wavelength, and Mie, when the particles are similar or bigger than the wavelength. Although the scattering can impact the lasercom link directly as an attenuation of the transmitted signal (see section 8.2.1), a potentially bigger impact is due to an indirect effect, i.e. the dispersion of light sources different to the communication laser, e.g. the Sun, the Moon, etc. This diffracted light can enter the field of view of the receiver, even if its orientation is angularly far away from these other light sources. Rayleigh scattering is the main contribution when the receiver is oriented with angles far away from the source, and Mie scattering prevails when the receiver points close to the source.

The radiation originated by the scattering of light different from the source is called sky radiance. When it enters the receiver's field of view, it adds to the communication signal as background noise, reducing the dynamic range of the system. This noise source can be modelled as a noise power $N_S$ (expressed in W), which is defined by the equation (8.6), where $L(\lambda, \theta, \varphi)$ represents the sky spectral radiance per area. $L(\lambda, \theta, \varphi)$ depends on the wavelength $\lambda$, the receiver's zenith angle $\theta$ and the angle $\varphi$ between the receiver and the Sun (or other noise source) and it is expressed in W/(cm²·srad·µm). For a given spectral radiance, the noise power $N_S$ depends on the receiver's aperture area $A_R$ (in cm²), the field of view $\Omega_{FOV}$ (in srad) and the filter's spectral width $\Delta\lambda$ (in µm).

$$N_S = L(\lambda, \theta, \varphi) A_R \Omega_{FOV} \Delta\lambda \qquad (8.6)$$

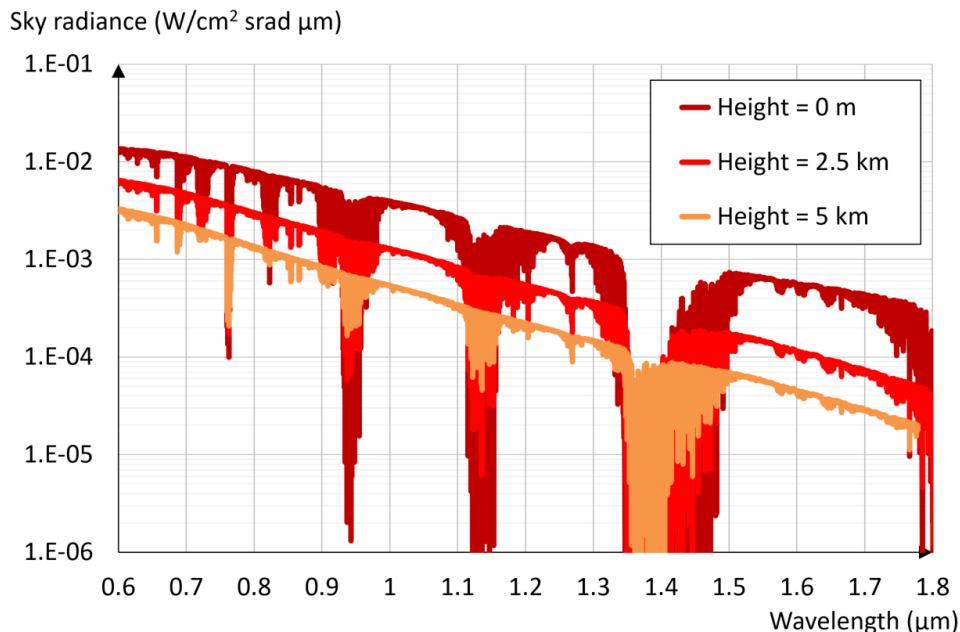

Fig. 8.5. Sky radiance received at three different heights at a zenith angle of 40° with a Sun zenith angle of 60° as a function of the wavelength with good visibility (23 km) and no clouds.

The sky radiance varies in a very wide range, depending on many factors like the presence of aerosols in the atmosphere or the sun angular position with respect to the link direction. The Fig. 8.5 shows an example of its dependence with the wavelength for three different heights of the lasercom-terminal location assuming an aerosols' rural model with good visibility (23 km) and no clouds. Between 0.8 and 1.5 µm, the most usual wavelengths in free-space, there is a difference of more than one order of magnitude, and another order of magnitude between sea level and 5 km at 1.5 µm (half an order for 0.8 µm). Other important factor determining the sky radiance is the angle between the receiver and the Sun. This angle is 10° in Fig. 8.5, although much smaller angles have already been demonstrated (e.g. NASA's LLCD mission went as close as 3° [9]). Fig. 8.6 shows this dependence, where the sky radiance is represented at 1.5 µm as a function of the zenith angle for several Sun zenith angles, assuming the same conditions of Fig. 8.5 and the middle-point height (2.4 km, where the astronomical observatories of Canary Islands, Spain, are located). A difference of more than two orders of magnitude in the sky radiance can be observed between the minimum and maximum angular separation.

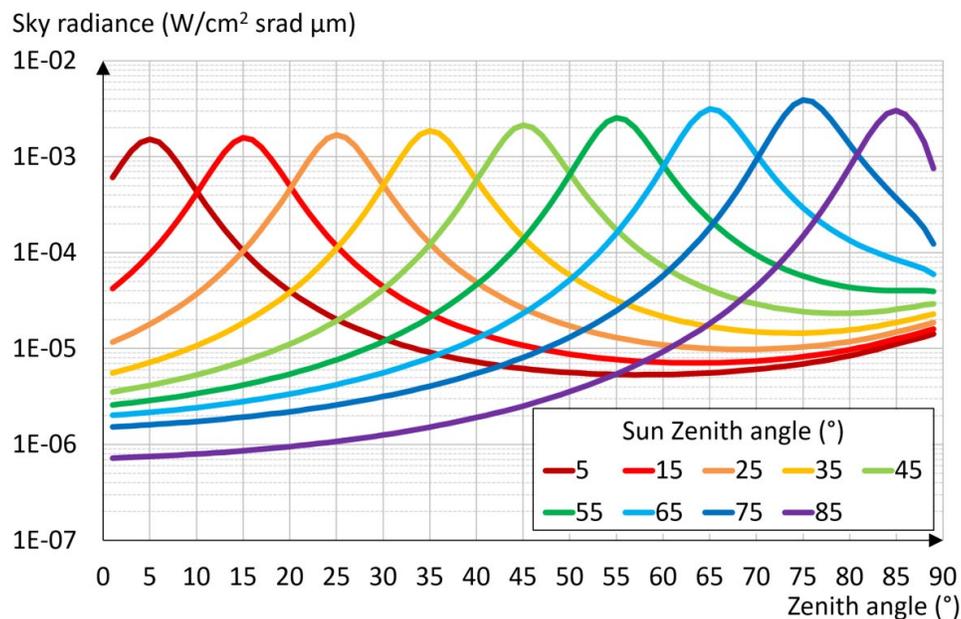

Fig. 8.6. Sky radiance as a function of the zenith angle for several Sun zenith angles at 1.5 µm in a height of 2.4 km for a rural model of aerosols with good visibility (23 km) and no clouds.

Analyzing the equation (8.6), the strategies to reduce the background noise due to the scattering of Sun's light can be deduced. On one hand, the sky radiance $L(\lambda, \theta, \varphi)$ is usually determined by the mission: the operating wavelength, the satellite's orbit, how close to the Sun the ground terminal can operate, and the time of the day when the operation is required (this last parameter being almost negligible during the night and a potentially-strong source of noise during the day). Reducing the receiver's aperture $A_R$ is not a good strategy to improve the SNR since it impacts negatively on the signal power. Spectral filtering to reduce $\Delta\lambda$ is an important technique, allowing to decrease the out-of-band noise, but not in the communication wavelength. Therefore, the field of view $\Omega_{FOV}$ would be the parameter to focus on in order to increase the SNR.

Field of view

The field of view (FOV) describes the angular extent that the object plane shows in the image plane of an optical system. FOV depends not only on the characteristics of the optical system, but also on the photodetector that captures the light of that system. Fig. 8.7 shows the FOV $\theta_{FOV}$ of a generic optical system characterized by its equivalent focal length $f$ and the size $d$ of a photodetector in the image plane.

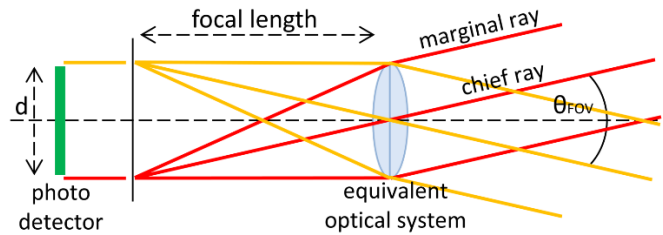

Fig. 8.7. Field of view of a generic optical-communication system defined by its focal length $f$ and a photodetector of size $d$.

Fig. 8.7 represents the convergence of two collimated beams on the equivalent focal plane of the receiving system, describing the widest angle for a given photodetector size $d$ and a focal length $f$. The chief ray (going through the center of the optical system) and the marginal ray (going through the edges of the aperture) describe completely the collimated beams going through an optical system. The FOV $\theta_{FOV}$ assuming a circular detector can be deduced by the equation (8.7). This equation shows that the FOV is proportional to the detector size and inversely proportional to the focal length, being determined by the chief ray. Therefore, according to the equation (8.6), if the lasercom system has to operate during the day under strong sky radiance, a small detector size and long focal length should be considered (which requires a better pointing accuracy) in order to minimize the background noise.

$$\theta_{FOV} = 2 \operatorname{atan}\left(\frac{d}{2f}\right) \qquad (8.7)$$

Modulation

The simplest way to modulate an optical signal consists in turning the transmitter's laser on and off (OOK, On-Off Keying), as in Fig. 8.8. This is an intensity binary-level modulation that allows to use direct detection, which is the most common technique due to its simplicity: IM/DD (Intensity Modulation/Direct Detection). These receivers convert the optical signal directly to an electrical current by using detectors following the square law, meaning that the electric output is proportional to the square of the amplitude of the electric field $E^2$ recovering directly the original intensity-modulated signal. This modulation allows relatively-high speed while keeping a low implementation cost, thus finding a good application in scenarios such as LEO-to-ground, especially when small satellites are required. Despite its simplicity, this scheme has relatively poor energy and spectral efficiency.

Pulse Position Modulation (PPM) is a variation of OOK, with less spectral efficiency but much more energy efficient. It finds a good application when the spacecraft energy resources are scarce and the losses are high, i.e. low photon flux links, such as deep space. Together with single-photon receivers, PPM is the optimum solution for photon-starved channels with data rates under ~Gbit/s. This modulation allows to encode more than one bit per pulse by dividing the duration of each sequence of n bits into m = 2n slots, corresponding to m symbols. When each pulse is sent, it is placed in one of the slots, defining the symbol to transmit (see Fig. 8.8, below). In this way, the duty cycle of the laser is reduced, transmitting a higher peak power for the same average power (the peak-to-average power is equals to m/n, compared to ½ with OOK) and improving the link SNR at the cost of a 1/m times lower spectral efficiency, requiring a higher modulation speed to keep the same binary rate. It is important to note that despite the large bandwidth available in lasercom compared to RF, the spectral efficiency can be a real limitation as well because it determines the speed of the technology needed to perform the modulation, which can become a hard requirement in high PPM orders. PPM can be considered as an encoded version of OOK, being based on IM/DD as well, finding its most efficient version when it is applied with photon-counting receivers. However, photon-counting receivers have a higher hardware complexity, being the synchronization of the received pulses the main limitation of high modulation orders in PPM.

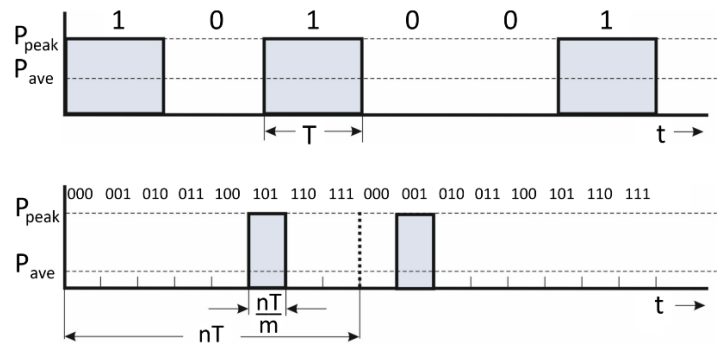

Fig. 8.8. Modulation of the sequence 101001 in OOK (above) and in 8-PPM (below).

Differential Phase-Shift Keying (DPSK) is another scheme that allows using IM/DD receivers based on a delay-line interferometer while showing a better background rejection than intensity modulations such as OOK or PPM, since the background noise is mostly added to the intensity of the signal, not to the phase. The demodulation is achieved by comparing the phase of two consecutive bits after splitting the incoming PSK signal into two separate paths with a time delay corresponding to one bit between them. Both signals can then be detected by a balanced receiver. DPSK is more robust against the atmospheric effects than intensity modulations, while being much simpler to implement than using a coherent phase receiver, and much more bandwidth-efficient than PPM. It is a good scheme to achieve high data rate while operating with the atmosphere as the channel.

Coherent demodulation consists in combining the received signal with a local oscillator in the optical domain so that the surface of the photodiode receives a mixture of both signals. For a coherent detector to work properly, it is essential that the local laser matches in frequency and phase to the received signal. When that condition is met, this scheme improves background-noise rejection because the received signal is amplified after mixing it with the local oscillator, resulting in a higher SNR. Equation (8.8) shows the relation between the SNR of a direct-detection receiver $SNR_{DD}$ and a coherent one $SNR_{CD}$, both being based on avalanche photodiodes (APD). $P_L$ and $P_R$ represent the local-laser power and the received-signal power respectively, and $M$, $x$, $R_0$, $I_d$ and $N_T$ refer to APD usual parameters, i.e., APD multiplication factor, dependence on the material, responsivity, darkness current, and spectral density of power of the thermal noise respectively. Equation (8.8) proves that if $P_L$ is big enough, the predominant noise is the shot noise, and $SNR_{CD}$ will always be bigger than $SNR_{DD}$.

$$\frac{SNR_{CD}}{SNR_{DD}} = \left(4\frac{P_L}{P_R}\right)\left(\frac{eM^{2+x}[R_0 P_R + I_d] + N_T}{eM^{2+x}[R_0 P_L + I_d] + N_T}\right) \quad (8.8)$$

Currently, coherent detection based on analogue optical phased-lock-loop is applied operationally in inter-satellite links. The high sensitivity of this reception technology allows transmitting over large distances (LEO to GEO). For satellite to ground links, adaptive optics (see section 8.4.4) at the receiver is required for either achieving a high-heterodyne efficiency when mixing the received signal with the local oscillator or for coupling into a single-mode fiber. Both approaches have similar requirements and depend on the relation between the receiver aperture diameter and the atmospheric coherent length (see section 8.2.2), i.e. the Fried parameter [15]. Experiments using coherent systems through the turbulence atmosphere have been performed, using both an analogue optical phase-locked loop [16] and digital signal processing [17]. However, for satellite-to-ground communications links up to 10 Gbit/s direct detection is preferred because of the reduced hardware complexity.

Receiver's sensitivity

Receiver's sensitivity is defined as the minimum power that a given system needs to reach a given quality measure. It is a fundamental parameter in every lasercom system because it determines all the other design choices. For any signal of a given average power, the photon arrivals in the receiver are not homogeneously distributed in time. Instead, the probability $P(n)$ of detecting $n$ photons during a certain period of time is the average number of photons $\mu$, which follows a Poisson distribution according to the equation (8.9).

$$P(n) = \frac{\mu^n}{n!} e^{-\mu} \qquad (8.9)$$

Considering an ideal detector where there is no noise in the '0's and the quantum efficiency is equal to 1, the decision threshold could be set at 0 because when no photon is transmitted, no photon can be detected. Therefore, the only source of error would be when a '1' is detected as a '0' because of the inhomogeneous Poisson distribution of photon arrivals. If '1's and '0's are equally distributed in the signal, the error probability $P_e$, or the probability of a '1' being detected as a '0' is $P(n = 0) = e^{-\mu}/2$, thus it is possible to derive the average number of photons per '1' bit $\mu$ for a given error probability $P_e$, or bit error rate (BER), which is given by equation (8.10) and represented in Fig. 8.9.

$$\mu = \ln\left(\frac{1}{2P_e}\right) \qquad (8.10)$$

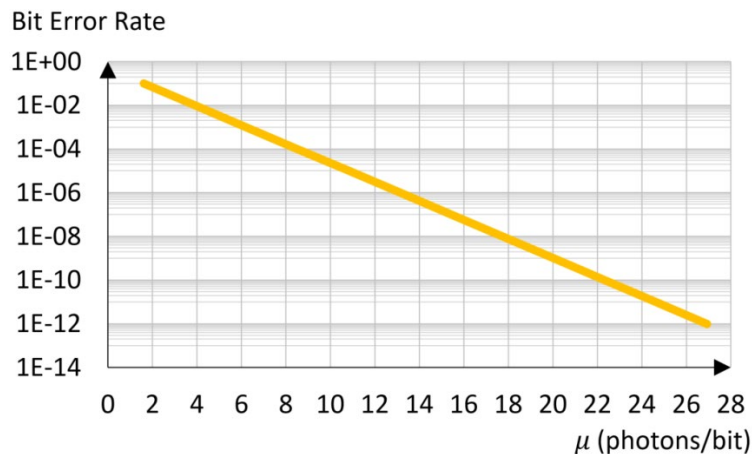

Fig. 8.9. Bit Error Rate as a function of the number of photons per bit.

The average received power $P_R$ can be calculated as the product of the photon arrival rate $N$ and the photon energy $h\nu$, where $h$ is the Planck's constant and $\nu$ is the frequency of the signal (equals to $c/\lambda$), according to the equation (8.11).

$$P_R = N \cdot h\nu = N\left(\frac{hc}{\lambda}\right) \quad (8.11)$$

Therefore, the received power to achieve a given error probability $P_e$ at a data rate of $1/T_b$, being $T_b$ the bit period, can be calculated as shown in equation (8.12). Fig. 8.10 shows the representation of equation (8.12), with the relation between the BER and the sensitivity limited by quantum noise for several typical data rates at a 1.55-µm wavelength. In practice, the sensitivity will be higher than this fundamental limit depending on the actual implementation of the receiver, being the pre-amplified coherent receivers the configuration achieving the best sensitivity [18] in high-data rate communications and single-photon counting PPM with lower data rates limited by the response of single-photon detectors [19].

$$P_R = \mu \frac{T_b}{2}\left(\frac{hc}{\lambda}\right) = \ln\left(\frac{1}{2P_e}\right)\frac{T_b}{2}\left(\frac{hc}{\lambda}\right) \quad (8.12)$$

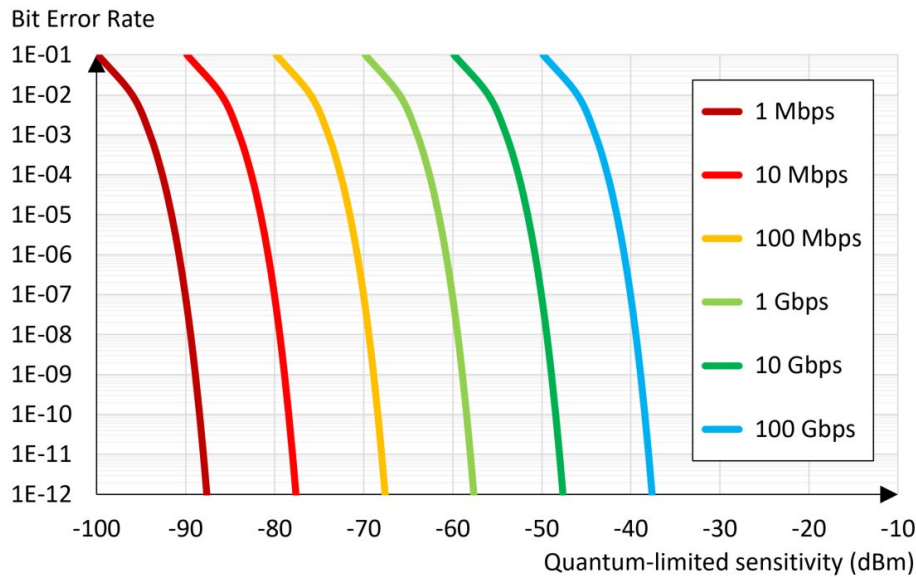

Fig. 8.10. Bit Error Rate as a function of the quantum-limited sensitivity.

Coding and interleaving

The previous section analyzed the quantum-limited performance of a noiseless system. In practice, for non-single-photon-PPM, when Additive White Gaussian Noise (AWGN) is present, the channel capacity $C$ is defined in bit/s by the Shannon-Hartley theorem in equation (8.13), where B is the available bandwidth in hertz and $P_R$ and $N_R$ are signal and noise power in watts. It gives another fundamental limit determining the relation between the maximum data rate that can be transmitted with an arbitrarily low BER applying the maximum possible efficiency of error-correcting coding [20].

$$C = B\log_2\left(1 + \frac{P_R}{N_R}\right) \quad (8.13)$$

The link data rate $R_b$ must be lower than the channel capacity C. Since $P_R = R_b E_b$, where $E_b$ is the energy per bit, and $N_R = N_0 B$, where $N_0/2$ is the noise variance, equation (8.13) can be expressed as in equation (8.14), where $E_b/N_0$ is called power efficiency and the $R_b/B$ is called bandwidth efficiency. Fig. 8.11 shows the plot of this equation, where the upper part symbolizes the area where error-free communication is not possible and the area where error-free communication is possible.

$$\frac{E_b}{N_0} > \frac{2^{R_b/B} - 1}{R_b/B} \qquad (8.14)$$

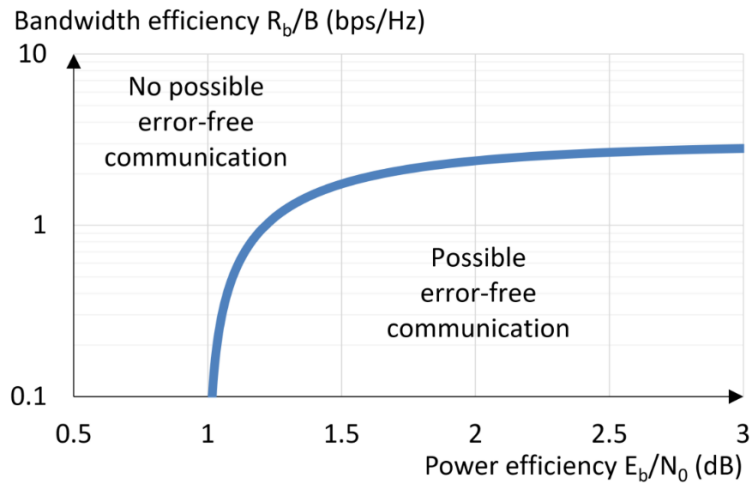

Fig. 8.11. Bandwidth efficiency vs power efficiency.

The first free-space lasercom links were based on getting the systems work more than building efficient systems. How the atmospheric effects impacted the communication performance was not well understood, and up to a certain point, it still is not. Therefore, the basic link-budget calculation was assumed adding a generous link margin to close the link successfully. More recently, the very well-known error-correcting coding techniques, used extensively in RF, have been applied to make lasercom efficient at the cost of some increase in the electronics complexity, but gaining several dB in the link margin and getting closer and closer to the theoretical channel capacity limits [21].

To apply a Forward Error Correction (FEC) coding successfully, the atmospheric-channel model is a key parameter, especially understanding how the intensity fluctuates due to the atmospheric turbulence, which is a big source of bit errors. There are many different models on the intensity fluctuation, although they are being gradually replaced by the gamma-gamma distribution [22] due to its good applicability to many turbulence regimes, from weak to strong, by multiplying two gamma intensity distributions which considers small and large-scale cells. When coding techniques are based on a realistic modeling of the channel, the SNR gain impacts directly in the reduction of the power and/or mass/size of the lasercom terminals.

All FEC codes add redundant bits to the output sequence to carry out the error correction in the received signals without retransmission while reducing the effective data rate depending on the expected errors in the channel, thus the amount of redundancy of the code. In the same way as in radiofrequency or optical fiber, the FEC codes used in free-space can be divided in block codes, such as Reed-Solomon (RS) or Low-Density Parity Check (LDPC), and convolutional codes, being the turbo codes the most efficient, performing very close to the Shannon limit at the cost of a higher decoding complexity.

Along with FEC codes, interleavers are usually employed in FSOC when the signals go through the atmosphere. Interleavers alter the sequential order of the transmitted bits to reduce the impact of burst errors by taking advantage of the fact that the channel fluctuates several orders of magnitude slower than the bit transmission time. This technique enhances the performance of FEC codes by uncorrelating the signal fading experienced by adjacent bits, distributing the errors in different code blocks in the case of block codes, or separating them in the case of convolutional codes. On the other hand, interleaving introduces latency in the communication and can require high-speed big memories to store the data during a typical fade, which is an expensive resource in space. As with FEC codes, there are block interleavers and convolutional interleavers, although the convolutional ones provide a reduction of 2× in both latency and memory requirements [23].

8.1.3 Link budget

The link budget is the key method to determine the overall performance of a lasercom system under a set of operating conditions. The basic link budget is given by equation (8.15) and relates the received power $P_R$ to the transmitted power $P_T$, the transmission and reception gain $G_T$ and $G_R$, the losses of the transmitter $L_T$ and the receiver $L_R$, the atmospheric losses $L_A$, the pointing losses $L_P$ and the free-space losses $L_S$.

$$P_R = P_T G_T L_T L_P L_S L_A L_R G_R \quad (8.15)$$

The most significant parameters in the link budget can be easily quantified, which allows making a quick preliminary analysis of the link. The transmitted gain $G_T$ and received gain $G_R$ can be calculated with equations (8.16) and (8.17), where $\Theta_T$ is the full transmitting divergence angle in radians, $D_R$ is the telescope aperture diameter, and $\lambda$ is the wavelength. The pointing loss $L_P$ is defined by the equation (8.18), where $\Delta_\Theta$ is the pointing accuracy. The free-space loss is given by equation (8.19), where $L$ is the distance between terminals.

$$G_T = \frac{16}{\Theta_T^2} \quad (8.16)$$

$$G_R = \left(\frac{\pi D_R}{\lambda}\right)^2 \quad (8.17)$$

$$L_P = exp\left(-2\left(\frac{\Delta_\Theta}{\Theta_T}\right)^2\right) \quad (8.18)$$

$$L_S = \left(\frac{\lambda}{4\pi L}\right)^2 \quad (8.19)$$

Table 8.1 and Fig. 8.13 show an example of a basic link-budget calculation for the LEO-to-ground SOTA mission carried out by NICT (Japan) [11]. The conditions of this link budget are as follows: the telescope's elevation is 30° for a link distance of 1,107 km between the ~600-km SOTA orbit and the NICT's OGS in Koganei (Tokyo, Japan) during the pass on December 9[th], 2015; the operating wavelength is 1549 nm; the receiver's aperture is 1 m; the optical signal is coupled into multi-mode optical fiber; and the transmitter, receiver and pointing losses are based on experimental measurements. As a reference, the received power measured in an experiment with the same conditions as in the link-budget calculation was -51.30 dBm. Since there are many factors affecting any link-budget calculation, a common practice is taking a few dBs as link margin.

Table 8.1. Link-budget calculation for LEO-to-ground mission.

| | |
|---|---|
| Transmitted power $P_T$ (dBm) | 15.40 |
| Transmitting gain $G_T$ (dB) | 85.08 |
| Transmitter loss $L_T$ (dB) | 1.97 |
| Pointing loss $L_P$ (dB) | 5.70 |
| Free-space loss $L_S$ (dB) | 259.06 |
| Atmospheric loss $L_A$ (dB) | 2.66 |
| Receiving gain $G_R$ (dB) | 126.14 |
| Receiver loss $L_R$ (dB) | 7.40 |
| Received power $P_R$ (dBm) | -50.18 |

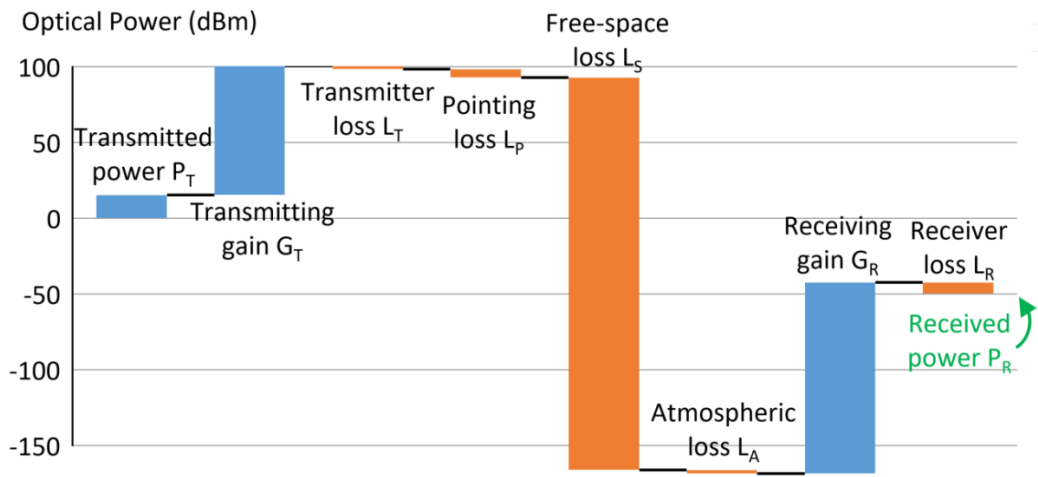

Fig. 8.12. Link budget for LEO-to-ground mission.

Operating wavelength

Looking at the equations (8.16), (8.17) and (8.19), it is easy to understand how important the wavelength choice is in a lasercom link. According to these equations, the shorter the wavelength, the higher the antenna gain and the lower the free-space losses are. Hence, from the geometrical point of view, shorter wavelengths are preferable (the same consideration is valid for increasing the transmitting and/or receiving aperture in terms of improving the delivery of power due to geometrical reasons: In the first case, the beam divergence gets reduced, and in the second case, more signal can be gathered). However, the strength of intensity fluctuations due to atmospheric turbulence increases with $\lambda^{-7/6}$, the atmospheric attenuation increases with $\lambda^{-2}$, and the scattering attenuation and sky radiance increase with $\lambda^{-4}$. Therefore, if the signal must go through the atmosphere, shorter wavelengths provide a larger scintillation (defined in section 8.2) with a stronger impact of the sky radiance.

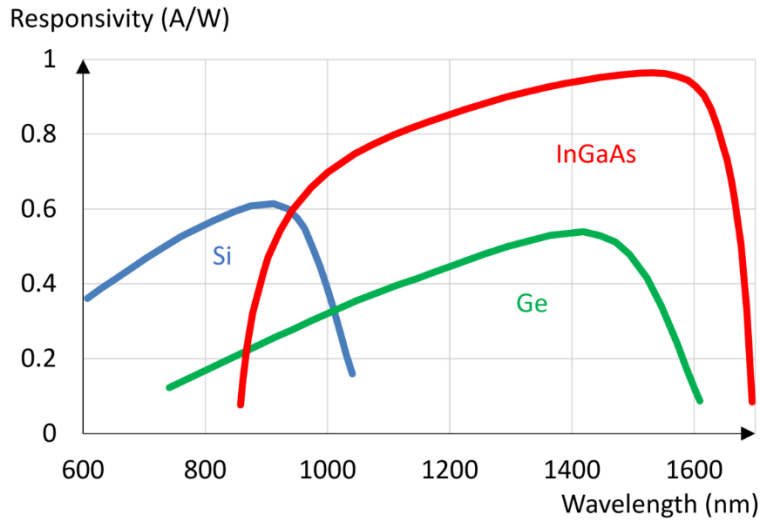

Fig. 8.13. Typical responsivity vs wavelength for silicon, InGaAs, and germanium.

Currently, there are three important regions where space lasercom operates. The most popular regions are around 1.064 µm and especially 1.55 µm if the laser beams propagate through the atmosphere because of their better behavior described in the previous paragraph. Furthermore, there is plenty of technology available at 1.55 µm from optical-fiber communications, the attenuation is lower and the responsivity of InGaAs photodetectors show a good behavior (see Fig. 8.13). From the eye-safety point-of-view, 1.55 µm is preferred because the eye fluids absorb these wavelengths before been focused on the retina. 1.064 µm has been developed mainly for inter-satellite links, where eye-safety is not an issue and there is no atmosphere in the channel. This shorter wavelength takes advantage of lower beam divergence and bigger antenna gain, while having available the Nd:YAG technology, especially suitable for coherent communication. When the receiver's noise is an issue, e.g. in quantum communications, wavelengths in the band of 800-900 nm can be preferred, taking advantage of the good responsivity of silicon photodetectors (see Fig. 8.13), which are less noisy than InGaAs and germanium.

8.2 Characteristics of the atmospheric channel

The atmosphere is transparent around the 800 nm, 1064 nm and 1550 nm, which allows using technologies based on Silicon, AlGaAs (Aluminium Gallium Arsenide), Germanium and InGaAs (Indium Gallium Arsenide). Most commercial-of-the-shelf (COTS) components for fiber communications are also usable, especially the ones at the optical C band, where there are less absorption peaks. High-precision lasers at 1064 nm are available also at high-power levels, making them suitable for coherent communications.

Apart from absorption or scattering effects, which can be taken into account by increasing a few dB the transmitted power (when not transmitting at absorption line of water or $CO_2$, for example), the atmospheric turbulence is the main challenge when transmitting through the atmosphere. Turbulence causes intensity fluctuations at the receiver which can seriously impair reliable communication. These fluctuations are due to self-interference processes and pointing jitter and they can be described statistically by a set of parameters, which are described in this section.

8.2.1 Attenuation: absorption and scattering

The optical signal in the atmosphere is attenuated by mainly two effects:

- Absorption: Caused by molecular absorption bands, molecular absorption continuums, and aerosols.
- Scattering: Caused by molecular effects (Rayleigh-Scattering) and by larger objects (Mie-Scattering), e.g. dust-particles or fog-drops.

Both, absorption and scattering are directly dependent on the particle density and the air density (molecule density). The Beer's law in equation (8.20) gives the power loss $L_a$ through the atmosphere for a given propagation distance $z$:

$$L_a = e^{-\alpha_e z} \text{ where } \alpha_e = \alpha_a + \alpha_{sc} \quad (8.20)$$

The extinction coefficient $\alpha_e$ (usually given in km$^{-1}$) is the sum of an absorption coefficient $\alpha_a$ and a scattering coefficient $\alpha_{sc}$ For sources with narrow spectral bandwidth such as lasers, the extinction coefficient $\alpha_e$ can be considered constant over the transmitted wavelength portion. The transmission spectrum under clear-sky conditions exhibits windows where atmospheric transmission is conceivable. The absorption coefficient $\alpha_a$ is highly wavelength-dependent: the atmospheric molecules absorb light whose wavelength matches the quantized energy levels of the molecules. The strength of these effects depends also on the location but it is mainly a function of altitude. There are several databases of scatter and absorption coefficients (for example, [24], [25]). Generally, the volcanic activity is taken into account since the concentration of aerosols in the atmosphere is highly dependent on it. But they do not include pollution, fog or other local effects, which should be added for each case.

The three common communication wavelengths (810, 1064 and 1550 nm) do not lie in spectrum lines of strong absorption. Besides, there is a decreasing slope as a function of wavelength: this corresponds to the scattering of the air particulates. A moderate volcanic activity has been assumed, which influences only high-altitude curves. The simple approximated model for the clear-sky attenuation in equation (8.21) can be used for a ground station at 100 m above the sea level [26].

$$\alpha_{atmos}(\varsigma) = \frac{10}{\left(\frac{\lambda}{1550nm}\right)^2 (\varsigma + 1)} [dB] \quad (8.21)$$

In Fig. 8.14, the clear-sky atmospheric attenuation is represented for both wavelengths, 1064 nm and 1550 nm. At 1064 nm, the atmospheric attenuation can reach 3 dB at low elevation angles, whereas the attenuation at 1550 nm is lower, remaining below 2 dB.

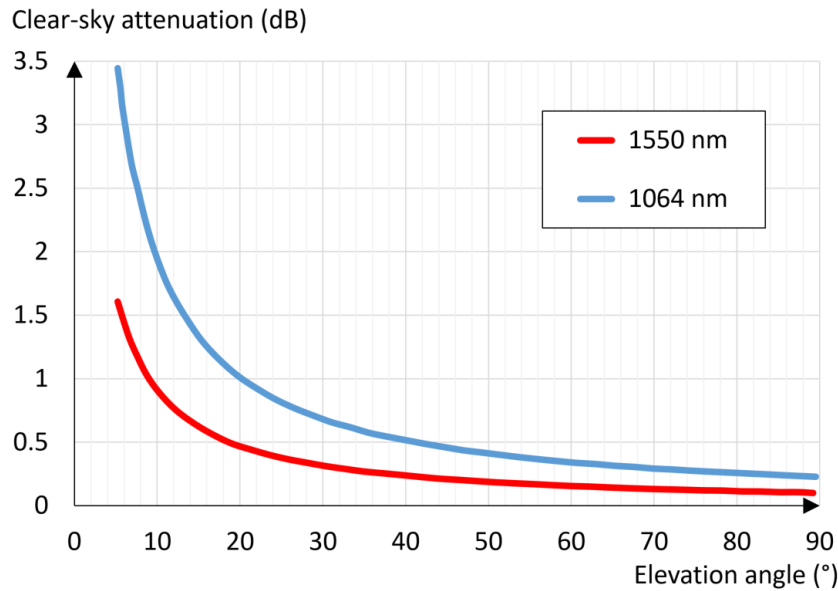

Fig. 8.14. Clear-sky attenuation versus elevation at 1064 nm and 1550 nm.

Attenuation caused by rain, snow, fog or clouds is however relatively independent of the optical wavelength. Fog and cloud attenuations are strong enough to make many optical links impossible. Near ground, thick fog can reach attenuation values of 300 dB/km ($\alpha_e$ = 69.8 km$^{-1}$) or more. In the troposphere, cloud attenuation values are on the order of 50 dB/km ($\alpha_e$ = 11.6 km$^{-1}$). Therefore, in presence of clouds optical links are usually not possible. To solve this issue, site diversity is required: a network of ground stations with uncorrelated atmospheric conditions to guarantee that at least one of them is available with a certain probability, increasing the overall network availability. The correlation of cloud coverage between two locations usually decreases for distances larger than 80 km. Considering ground stations at both hemispheres contributes to reach a higher availability with less number of stations. It is important to note that the availability of one single station is not as important as its decorrelation with the other stations in the network. Therefore, stations at locations with high cloud coverage probability can also contribute positively to the network availability.

8.2.2 Atmospheric turbulence

Atmospheric-turbulence modelling is a wide field, which is not pretended to be fully covered in this section. An extensive research on atmospheric turbulence and its effects is available in the literature, covering theory, simulations and measurements [27] - [28]. The objective of this section is to define the main parameters used to describe the turbulent effects relevant to the communications link.

From the communications perspective, the turbulence impact can be described with the following parameters:

- Scintillation: normalized variance of the signal fluctuations.
- Intensity correlation time: describes the time characteristics of the signal fading.
- Fading distribution: statistical distribution of the intensity fluctuations.

Turbulence creates phase-front distortions on the laser beam propagating through the atmosphere. These distortions lead to destructive and constructive interferences, redistributing the intensity within the beam. The statistical behavior of the atmospheric refractive index is used to model the turbulence and it can be expressed for optical wavelengths in terms of temperature and pressure. Since pressure fluctuations are normally negligible, the index of refraction follows mainly the fluctuations of the air temperature. Therefore, turbulence, i.e. refractive index fluctuations, is created mainly by the mixing of warm and cool air, either by convection or by wind shear. These two phenomena produce eddies (cells) of large scales that break up into smaller eddies, forming an energy transfer cascade. Turbulence cell sizes between a lower limit, called inner-scale, and an upper limit, called outer-scale, contribute to the optical turbulence. The range between these two sizes is called inertial range. Eddies, below inner scale, belong to the so-called viscous dissipation range, where the remaining energy in the fluid motion is dissipated as heat. The power spectral density of the refractive index is used to define the contribution of eddies to the atmospheric turbulence. Several models are used to describe the power spectral density considering either one or both inner and outer scales. However, Kolmogorov model is generally assumed, where inner scale is set to zero and outer scale to infinite [27]. The main reason is to keep the formulation relative simple and to avoid estimating inner and outer scales, which usually require local measurements, not easily practicable. The Kolmogorov model can provide a good estimation; however, it cannot describe all the turbulence effects.

Atmospheric turbulence is a random spatiotemporal field, assumed locally homogeneous and defined by variations of the index of refraction around its mean value. This last definition allows considering the statistical process stationary in increments, while assuming the mean to change slowly. The structure function is used for the statistical description. This second-order moment is defined as the difference between covariance functions at zero and after a stationary time increment. To relate both spatial and temporal domains, the "frozen turbulence" Taylor hypothesis is assumed. That means the temporal changes in the atmosphere depend only on the wind velocity orthogonal to the propagation path. The structure function of the refractive-index fluctuations is defined as a power law of the distance between two points. This function is directly proportional to a constant, defined as the structure parameter, which describes turbulence strength. The value of the structure parameter depends mainly on two things. First, on the altitude above ground: the atmosphere is denser close to the ground, which means that the structure parameter tends to decrease with the altitude. Second, on the time of the day: during sunset and sunrise, the atmospheric becomes quieter because the temperature gradient between the ground and the air decreases and during midday, on a sunny day, the structure parameter tends to be maximal.

For ground-to-space satellite communications, a profile of the refractive-index structure parameter is required to describe the turbulence strength along the transmission path. The turbulence strength is described by the structure parameter. The structure parameter is a multiplicative factor of the structure function, i.e. the covariance function assuming that the turbulence is statistically homogenous and isotropic. Averaged profiles are usually used to qualitatively describe the atmospheric impact. One of the most-widely used is the Hufnagel-Valley profile in the equation (8.22), which depends basically on the altitude above ground $h$, the turbulence on ground level, defined by the structure parameter at zero height $A = C_n^2(h = 0)$ and the mean cross-wind velocity $v$.

$$C_n^2(h) = Ae^{-h/100} + 2.7 \times 10^{-16}e^{-h/1500} + 0.00594(v/27)^2(10^{-5}h)^{10}e^{-h/1000} \quad (8.22)$$

For night time, these parameters are usually set to $A = 1.7 \times 10^{-14} m^{-2/3}$ and $v$ = 21 m/s and for daytime to $A = 1.7 \times 10^{-13}$ m$^{-2/3}$ and $v$ = 30 m/s. More complex profiles, with more input parameters may properly fit the turbulent path accurately but, as the number of variables increases, the need of an accurate estimation of such parameters may limit their practicability. Using the Hufnagel-Valley profile and the Kolmogorov power spectral density of the refractive index fluctuations, a first order estimation of the chosen parameters can be done.

Scintillation

Scintillation is defined as the variance of the signal normalized to its squared mean. This parameter is used to describe the temporal or spatial fluctuations of the received signal. The so-called scintillation index is used to describe the irradiance fluctuations of the optical wave in one single point, i.e. by a point receiver (a receiver can be defined as point receiver when its aperture is smaller than the intensity correlation length, i.e., it will be uniformly illuminated). When the receiver aperture increases and becomes bigger than the correlation length of the intensity fluctuations, scintillation decreases due to the so-called aperture averaging. That means the receiver can collect multiple correlation lengths across the transversal plane of the beam, averaging the signal fluctuations.

In satellite communications, the aperture size of the receiver, compared to the received intensity correlation length, differs between uplink and downlink. For the uplink, the satellite is a point receiver because the correlation length can be of several hundred meters, much larger that the telescope diameter. That happens because the turbulence is closer to the transmitter and after leaving the atmosphere only light diffraction happens. For the downlink, the receiver aperture is usually larger than the intensity correlation length, leading to aperture averaging and a reduction of the scintillation. For a point receiver, scintillation can be estimated with the equations (8.23) and (8.24).

$$\sigma_{I,downlink}^2 = \exp\left[\frac{0.49\sigma_{Bd}^2}{\left(1+1.11\sigma_{Bd}^{12/5}\right)^{7/6}} + \frac{0.51\sigma_{Bd}^2}{\left(1+0.69\sigma_{Bd}^{12/5}\right)^{5/6}}\right] - 1 \qquad (8.23)$$

$$\sigma_{I,uplink}^2 = \exp\left[\frac{0.49\sigma_{Bu}^2}{\left(1+0.56\sigma_{Bu}^{12/5}\right)^{7/6}} + \frac{0.51\sigma_{Bu}^2}{\left(1+0.69\sigma_{Bd}^{12/5}\right)^{5/6}}\right] - 1 \qquad (8.24)$$

where,

$$\sigma_{Bd}^2 = 2.25k^{7/6}\sec(\varsigma)^{11/6}\int_{h_0}^{H} C_n^2(h)(h-h_0)^{5/6}dh \qquad (8.25)$$

$$\sigma_{Bu}^2 = 2.25k^{7/6}L^{5/6}\int_{h_0}^{H} C_n^2(h)\left(1-\frac{h-h_0}{H-h_0}\right)^{5/6}\left(\frac{h-h_0}{H-h_0}\right)^{5/6}dh \qquad (8.26)$$

Both equations are valid under all turbulence regimes and give an estimation of the maximum scintillation that can be expected. The wavenumber is represented by $k = 2\pi/\lambda$, where $\lambda$ is the wavelength, the link distance is $L$ and the zenith angle of the link path is $\varsigma$. The profile is integrated between the OGS height above ground $h_0$ and the satellite height $H$.

Scintillation values after aperture averaging can be estimated by the equation (8.27), which is however only valid under weak turbulence conditions.

$$\sigma_{I,av}^2 = 8.70k^{7/6}(H-h_0)^{5/6}\sec(\varsigma)^{11/6}\Re\left\{\int_{h_0}^{H} C_n^2(h)\left[\left(\frac{kD^2}{16L}+i\frac{h-h_0}{H-h_0}\right)^{5/6}-\left(\frac{kD^2}{16L}\right)^{5/6}\right]dh\right\} \quad (8.27)$$

As a rule of thumb, when using the Hufnagel-Valley model for the vertical turbulence profile, the weak turbulence approximation holds its validity until 30° for low-strength turbulence. If the turbulence is stronger, the threshold can rise to 60°.

Measured power fluctuations at the satellite are the combination of scintillation and beam wander or pointing jitter. Assuming that the ground station is open-loop pointing towards the satellite, the root-mean-square (RMS) atmospheric-induced beam wander can be calculated as in equation (8.28).

$$\sigma_{BW} = 0.73\left(\frac{\lambda}{2W_0}\right)\left(\frac{2W_0}{r_0}\right)^{5/6} \quad (8.28)$$

When the point-ahead angle towards the satellite is small enough, within the coherence cone of the atmosphere, the angle-of-arrival fluctuations measured in the downlink can be used to pre-correct the beam wander. This point is discussed in a following subsection.

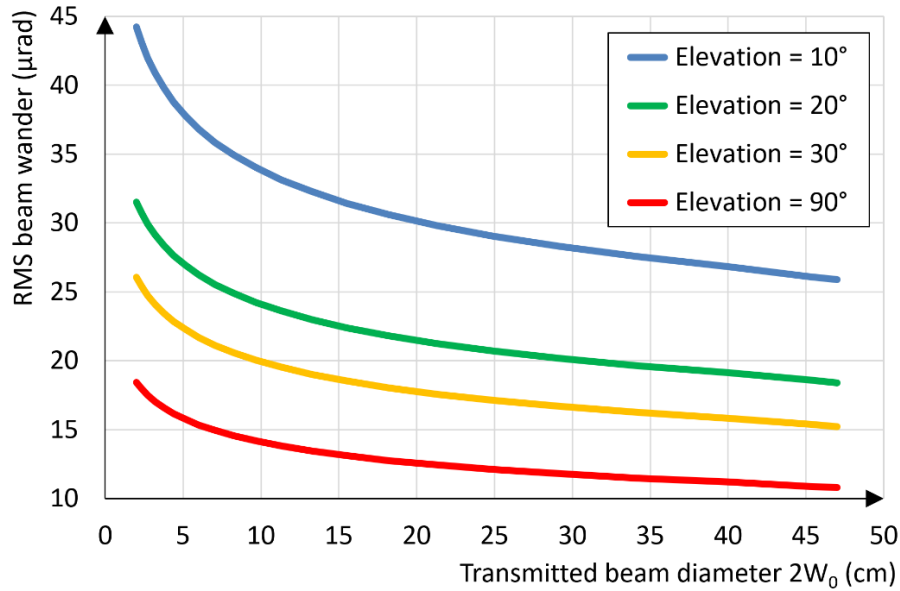

Fig. 8.15. RMS beam wander versus transmitted beam diameter for several elevation angles.

In Fig. 8.15, the RMS beam wander is plotted versus the transmitted beam diameter for several elevation angles. The RMS beam wander decreases when increasing the transmitted beam diameter, but its impact is larger, as it is shown in Fig. 8.17.

Correlation time and pointing jitter

The atmospheric wind impacts on the temporal bandwidth of the phase perturbations and it can be modelled by a Gaussian (also called Bufton) model, which is given by equation (8.29).

$$V(h) = \omega_s[L - L(h)] + 20 + 30\exp\left(-\left(\frac{h-9400}{4800}\right)^2\right) \quad (8.29)$$

The wind profile was modified, adding the first term, which takes into account the slew rate $\omega_s$ of the satellite with respect to the ground station; i.e. the angular velocity of the satellite seen by the ground station. This is an important factor especially for LEO satellite links, because of the satellite motion, which virtually contributes as cross-wind.

The distance between the ground station and any point in the line-of-sight to the satellite $L(h)$ can be calculated taking the Earth curvature $R_e$ and the link elevation angle into account as in equation (8.30), where the link distance can be calculated as $L = L(h = H_{satellite})$.

$$L(h) = -(R_e + H)cos(\varsigma) + [(R_e + H)^2[cos(\varsigma)]^2 + h^2 - H^2 + 2(h - H)R_e]^{1/2} \tag{8.30}$$

The correlation time in the signal fluctuations is defined by the greenwood frequency, i.e. the characteristic frequency of the atmosphere, that can be estimated by equation (8.31) [29].

$$f_G = 2.31\lambda^{-6/5}\sin^{-3/5}(\varsigma)\left[\int_{h_0}^{H} C_n^2(h)V^{\frac{5}{3}}(h)dh\right]^{3/5} \tag{8.31}$$

The correlation time can be then calculated by its inverse $\tau_c = 1/f_G$ and it describes the time behavior of the intensity fluctuations. As ccintillation is a self-interference process induced by the phase distortions produced by the atmospheric turbulence, $\tau_c$ is introduced to describe also the temporal behavior of the phase distortions. This parameter is especially interesting when designing FEC coding. In Fig. 8.16, the Greenwood frequency is shown versus the link elevation angle. In the calculation, the virtual wind due to the satellite movement has been taken into account, by setting the corresponding slew rate $\omega_s$, as defined in equation (8.29).

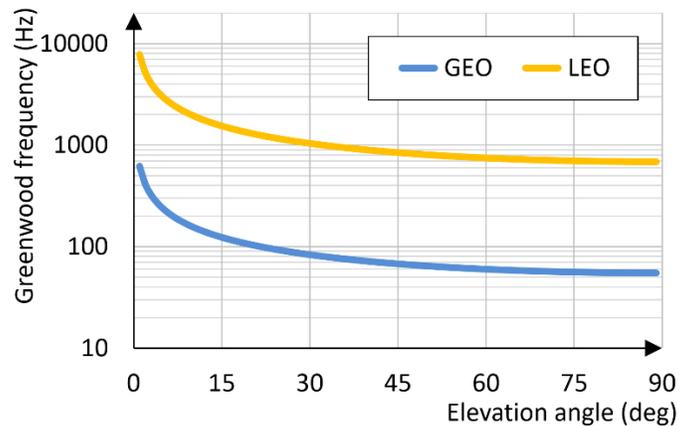

Fig. 8.16. Greenwood frequency $f_G$ versus elevation angle.

In free-space optical communications, the tracking of the optical signal is fundamental to guarantee a stable link. The tilt component of the phase, i.e. the angle of arrival, has another time characteristic as the one considering the whole phase distortions and it can be calculated with the equation (8.32) [30].

$$f_{TG} = 0.331\frac{1}{D^{1/6}\lambda\sqrt{cos(\varsigma)}}\left[\int_{h_0}^{H} C_n^2(h)V^2(h)dh\right]^{1/2} \tag{8.32}$$

This parameter is especially interesting for designing the tracking system. Such systems work in closed loop and the residual error due to the limited closed-loop bandwidth will produce a jitter. The variance of this angular jitter can be calculated by equation (8.33), where the transfer function $H(f) = 1/(1 + i f/f_{3dB})$ of the closed loop is assumed.

$$\sigma_{TG}^2 = \left(\frac{f_{TG}}{f_{3dB}}\right)^2 \left(\frac{\lambda}{D}\right)^2 \quad (8.33)$$

The equation (8.33) refers only to one-axis jitter, which is assumed to be normal distributed. Assuming that both axes are independent, the statistical distribution of the pointing error should be Rayleigh distributed, with its variance given by equation (8.34).

$$\sigma_{jitter}^2 = \left(\frac{4-\pi}{2}\right)\sigma_{TG}^2 \quad (8.34)$$

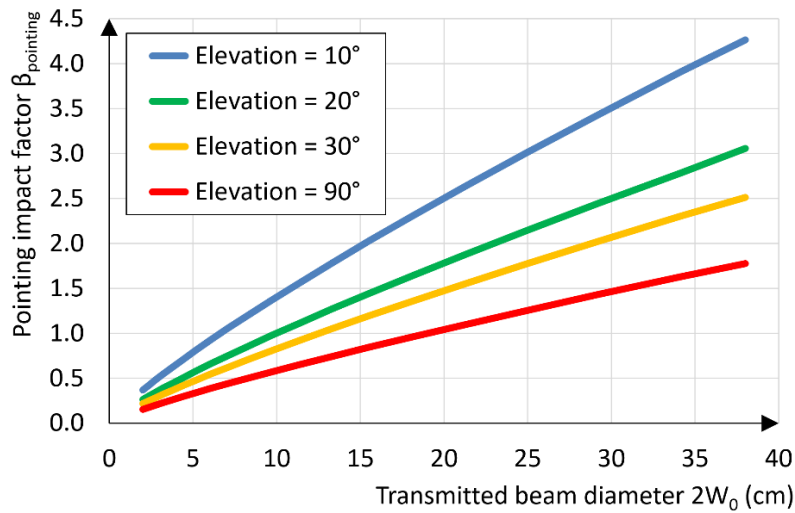

Fig. 8.17. Pointing impact factor $\beta_{pointing}$ vs transmitted beam diameter for several link elevation angles.

For GEO satellites, the uplink pointing uses the information of the downlink angle-of-arrival fluctuations to point towards the satellite and pre-correct the pointing fluctuations of the reciprocal path. Therefore, assuming that transmitter and received beam travels the same path and are driven by the same tracking-pointing mirror, the jitter in the tracking system is translated to the pointing system. For LEO satellites, this approach is not used because the atmospheric path of the uplink and downlink are completely uncorrelated due to the point-ahead angle required to hit the satellite.

The residual pointing jitter impact in the communication can be estimated by normalizing the $\sigma_{pointing}^2$ by the beam divergence $\vartheta_{beam}$, here defined as radius of the transversal Gaussian wave, when the intensity decays to the $1/e^2$. The impact factor $\beta_{pointing}$ is defined then by equation (8.35).

$$\beta_{pointing} = \frac{\sqrt{\sigma_{pointing}^2}}{\vartheta_{beam}}. \quad (8.35)$$

Pointing errors can be usually assumed negligible when $\beta_{pointing}$< 0.2. Here, $\beta_{pointing}$ represents how big is the pointing error with respect to the beam size. When $\beta_{pointing}$ = 1, it means that the shift due to the pointing error is equal to the beam size. However, one may perform an optimization of the beam-wander impact (signal fluctuations) with respect to the mean received power. Setting a beam divergence ten times larger than the beam wander implies a huge penalty in terms of received power.

Therefore, by designing an appropriate coding to protect data transmission, it is possible to reduce substantially the beam divergence. In Fig. 8.17, the $\beta_{pointing}$ is plotted versus the transmitted beam diameter, which is directly related to the beam divergence as $\vartheta_{beam} = \lambda/\pi W_0$. For larger beam diameters (narrower beam divergence), the impact on the pointing is larger, although the RMS beam wander is smaller, as already shown in Fig. 8.15. In this calculation, $\lambda$ = 1550 nm and the Hufnagel-Valley profile for the structure parameter.

Fading distribution

Due to intensity fluctuations, the power might decrease and assume values below the sensitivity threshold of the receiver, leading to a fading or loss of communication. Evaluating the probability of falling under a determined threshold will tell how much the power at the transmitter should be increased in order to ensure the communication availability stays within a desired probability, the so-called fading level.

Fading can be modelled assuming a log-normal probability density function (PDF), especially in weak turbulence conditions. The probability that the received power falls below the threshold $P_{min}$ is defined as in the equation (8.36), where $P_{min}$ is the receiver sensitivity, $P_0$ is the averaged received power and $\sigma_P^2$ is the scintillation of the received signal.

$$p_{thr}(P_{RX} < P_{min}) = \frac{1}{2}\left(1 + erf\left\{\frac{ln\left[\frac{P_{min}}{P_0}(\sigma_P^2 + 1)^{1/2}\right]}{2ln(\sigma_P^2 + 1)^{1/2}}\right\}\right) \quad (8.36)$$

Fade level is considered in the link-budget calculation as an extra-loss $\alpha_{scint}$, defined by equation (8.37).

$$\alpha_{scint} = 10\log_{10}\left(\frac{P_{min}}{P_0}\right) \quad (8.37)$$

Equation (8.38) combines both equations (8.36) and (8.37).

$$\alpha_{scint} = 4.343\left\{erf^{-1}(2p_{thr} - 1)[2ln(\sigma_P^2 + 1)]^{1/2} - \frac{1}{2}ln(\sigma_P^2 + 1)\right\} \quad (8.38)$$

Under strong turbulence, the intensity distribution does not follow a lognormal distribution. Gamma-gamma and Weibull [31] distribution can be used instead, for all turbulence conditions. However, a lognormal assumption can provide a conservative assumption in most of cases [32].

The lognormal distribution can be modified to include the effects of the pointing jitter as proposed in [33]. In this case, it is assumed that the pointing jitter and the scintillation are two independent random processes and therefore PDFs can be multiplied. The lognormal distribution is used for the scintillation and the beta-distribution is used for the pointing jitter. The pointing jitter is assumed Rayleigh distributed (normal in each single axis) and it is applied to Gaussian beam form to calculate the intensity fluctuations, leading to a beta-distribution. In this case, the combination of both PDFs is written the equation (8.39), where $I/\bar{I}$ is the normalized intensity, $\sigma_I^2$ is the scintillation, and $\beta = \theta_{div}^2/(8ln2)\sigma_{jitter}^2$. $\theta_{div}^2$ and $\sigma_{jitter}^2$ are respectively the equivalent beam divergence and pointing jitter at the far-field [34].

$$p(I) = \frac{1}{2\bar{I}}erfc\left[\frac{ln(I/\bar{I}) + \sigma_I^2(\beta + 1/2)}{\sqrt{2}\sigma_I}\right] \times \beta exp\left[\frac{\sigma_I^2}{2}\beta(\beta + 1)\right]\left(\frac{I}{\bar{I}}\right)^{\beta-1} \quad (8.39)$$

8.3 Low-Earth-orbit satellite communications

Low-Earth Orbit (LEO) offers several important advantages to spacecraft such as a benign radiation environment, a close distance to Earth, a low communication latency, and frequent and non-expensive launches. For these reasons, LEO is the most usual scenario for small satellites in general [35] and the one with the fastest expected growth. The lower cost compared to other orbits makes LEO suitable for missions consisting in single experiments as well as for deploying many satellites in constellations, where they can provide a unique coverage of the entire Earth. Launching a satellite to an orbit beyond LEO makes it more difficult to comply with the 25-year post-mission lifetime guideline set by the IADC (Inter-Agency Space Debris Coordination Committee) in order not to become space debris. For example, CubeSats in orbits above 750 km take centuries to decay, thus requiring some drag strategy. As will be described below, the most important characteristics that define lasercom from LEO are on one hand the need of performing accurate pointing and tracking to keep the link between the fast-moving satellite and the ground station, and on the other hand the propagation of the laser signals through the turbulent atmosphere before reaching the receiver on the ground. It will be shown that the basic strategies to overcome the challenges of this type of lasercom links have been already defined with minor differences, which makes it possible to describe the fundamental principles of this application.

8.3.1 Heritage

The first successful bidirectional LEO-ground lasercom experiment was carried out by JAXA and NICT using LUCE (Laser Utilizing Communications Equipment) onboard OICETS (Optical Interorbit Communications Engineering Test Satellite) in 2006 [36], which was a 570-kg satellite inserted in a 610-km LEO orbit. LUCE was a 100-kg lasercom terminal based on a 2-axis gimballed 26-cm Cassegrain telescope transmitting a 100-mW 847-nm laser at 50 Mbit/s, and an accurate fine-pointing system to control a beam with a footprint as small as 5 m reaching the ground station, where a 20-cm telescope was used to gather the received signal coupling it into an APD.

In 2010, the Department of Defense of the USA launched the NFIRE (Near Field Infrared Experiment) LEO satellite with a Tesat's LCT (Laser Communications Terminal) onboard. Although the goal of this terminal was to carry out inter-satellite links, it was used for LEO-to-ground links as a demonstration. LCT was made up by a 2-axis gimballed mirror assembly before a fixed 125-mm telescope, transmitting a 1-W 1064-nm laser at 5.6 Gbit/s using homodyne BPSK with no beacon. Because of using coherent detection on the ground, this system could only close the communication link by using a small receiver's aperture (6 cm) located in astronomical observatories in order to watch a Fried parameter larger than the receiver's aperture [37].

China launched in 2011 its first lasercom payload LCE (Laser Communication Equipment) to a 971-km LEO orbit onboard the Haiyang-2A (or HY-2A) ~1,500-kg satellite. The lasercom system was based on a 15-cm gimballed telescope and could transmit its 1-W laser beam with a tracking accuracy in the order of 1 μrad achieving a maximum data rate of 504 Mbit/s [38].

In 2014, JPL installed the OPALS (Optical PAyload for Lasercom Science) terminal in the International Space Station (ISS), at a 408-km LEO orbit [39]. OPALS used a 2-axis gimbal to move a 5-cm telescope and transmit a 2.5-W laser at 50 Mbit/s. The generous transmitted power allowed to relax the pointing accuracy down to 300 µrad, enough to point the ~1-mrad-divergence beam to the ground station.

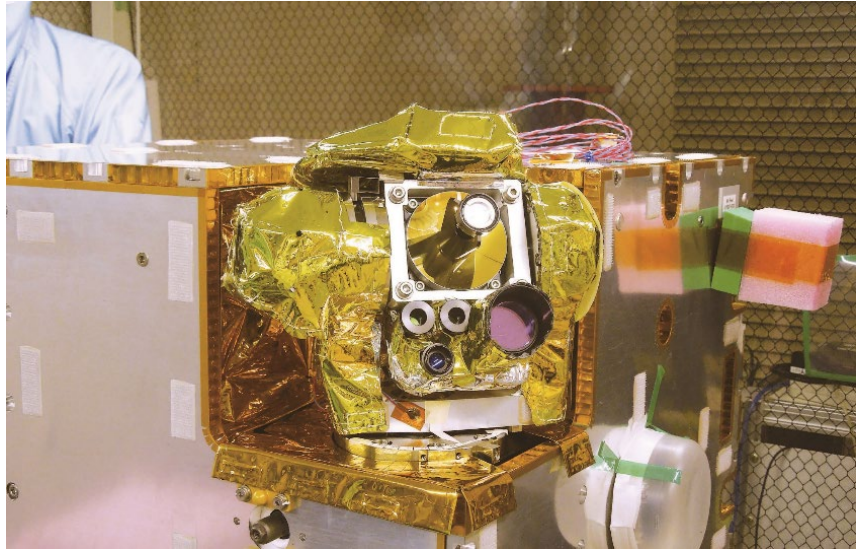

Fig. 8.18. Flight model of NICT's SOTA onboard SOCRATES.

The first lasercom system onboard a small satellite was NICT's SOTA (Small Optical TrAnsponder) [11] onboard SOCRATES (see Fig. 8.18), which was launched in May 2014 into a 628-km LEO orbit. SOTA was a 2-axis gimballed terminal with capabilities to perform a variety of lasercom experiments in a less-than-6 kg compact package. The core experiment was the 10 Mbit/s links at 1549 nm using coarse and fine-pointing to accurately transmit the 35-mW laser through a 5-cm Cassegrain telescope. SOTA had other additional capabilities, i.e. B92-like QKD protocol at 800-nm band to perform the first-time quantum-limited demonstration from space [40], and 10 Mbit/s downlinks at 980-nm using a small lens, both based on coarse-pointing only. As a collaboration with the Tohoku University, NICT developed a simplified version of SOTA called VSOTA (Very Small Optical TrAnsponder) with a weight of less than 0.7 kg based on body pointing only, thus transmitting a 1550-nm laser beam with a wide divergence (1.3 mrad), low power (80 mW) at a low data rate (up to 1 Mbit/s) [41]. VSOTA onboard Hodoyoshi-2 (RISESat) lost its launch opportunity planned for 2013 and was successfully launched in January 2019. Based on the SOTA and VSOTA heritage, NICT is currently working towards the next generation of miniaturized high-speed lasercom transmitters compatible with CubeSat platforms for LEO-GEO intersatellite links as well as LEO-ground.

Table 8.2. Specifications of in-orbit LEO-to-ground lasercom terminals.

|  | LUCE | LCT | LCE | OPALS | SOTA | OSIRISv2 | OSIRISv1 | MCLCD |
|---|---|---|---|---|---|---|---|---|
| Satellite | OICETS (570 kg) | NFIRE (494 kg) | Haiyang-2A (1500 kg) | ISS (420 tons) | SOCRATES (48 kg) | BIROS (130 kg) | Flying laptop (120 kg) | Micius (631 kg) |
| Operator | JAXA, Japan | DoD-MDA, USA | SOA, China | NASA-JPL, USA | NICT, Japan | DLR, Germany | DLR, Germany | CAS, China |
| Launch date | Aug 23, 2005 | April 24, 2007 | Aug 16, 2011 | April 18, 2014 | May 24, 2014 | June 22, 2016 | July 14, 2017 | Aug 15, 2016 |
| LEO altitude | 610 km | 495 km | 971 km | 408 km | 628 km | ~500 km | ~600 km | ~500 km |
| Mass | 100 kg | 35 kg | 67.8 kg | <180 kg | 5.9 kg | 1.65 kg | 1.34 kg | - |
| Beacon | 808 nm | No beacon | - | 976 nm | 1 µm | 1.56 µm | No beacon | 532 nm |

|  | CW |  |  | CW | CW | modulated |  | CW |
|---|---|---|---|---|---|---|---|---|
| Downlink | 847 nm | 1064 nm | - | 1.55 µm | 800, 980, 1549 nm | 1.545, 1.55 µm | - | 1549.731 µm |
| Modulation | On-Off Keying | Homodyne BPSK | - | On-Off Keying | On-Off Keying | On-Off Keying | On-Off Keying | DPSK |
| Max. bitrate | 50 Mbit/s | 5.6 Gbit/s | 504 Mbit/s | 50 Mbit/s | 10 Mbit/s | 1 Gbit/s | 200 Mbit/s | 5.12 Gbit/s |

Since 2008, DLR is developing optical terminals for small satellites (from CubeSats to ~100-kg class) in 1550-nm wavelength based on COTS components to provide solutions to small satellites with limited mass and power. In 2018, the first two generations of terminals were launch to space, which base the pointing on the satellite attitude control, reducing the terminal mass to the 1-kg class. The same approach is used in a CubeSat terminal in collaboration with Tesat Spacecom with a 10×10×3 cm$^3$ form factor, 300-gr mass and 100-Mbit/s data rate with fine-pointing. The 5-kg class OSIRISv3 will include coarse-pointing to avoid changing of the satellite attitude, providing 10 Gbit/s [42]. The first generation OSIRISv1, was launched in July 2017, onboard the Flying Laptop satellite of University of Stuttgart. DLR's OSIRIS onboard BIROS, known as OSIRISv2 [43] was launched in June 2016 into a 500-km LEO orbit, including an InGaAs four-quadrant-tracking sensor to track the 1560-nm modulated beacon and close the loop with satellite attitude control. The terminal is designed for downlinks up to 1 Gbit/s using an OOK-modulated 1-W 1545-nm laser through a 1.5-cm lens with a 200-µrad divergence. OSIRISv2 also includes another downlink capability up to 150 Mbit/s using a separate 1.5 cm lens with a divergence of 1,200 µrad and a transmitted power of 150 mW at 1550 nm. Both terminals were under commissioning at the time of writing this chapter.

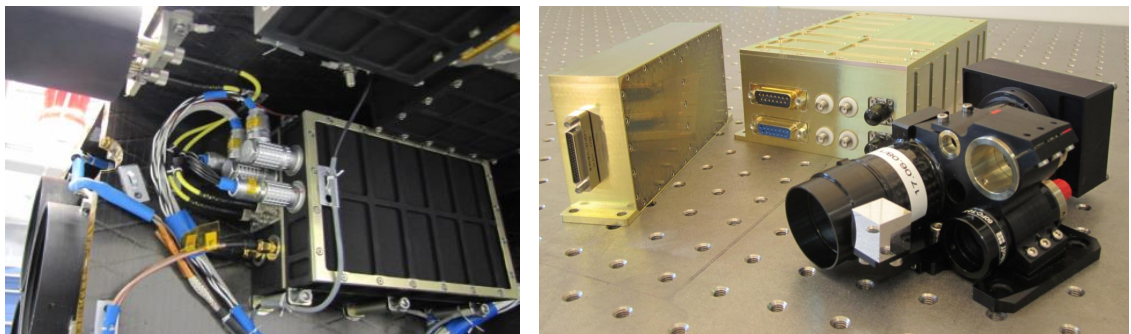

Fig. 8.19. OSIRISv1 onboard the Flying-Laptop satellite of University of Stuttgart (left). OSIRISv2 onboard the DLR's Biros Satellite (right).

In August 2016, the Chinese Academy of Sciences launched the Micius LEO satellite to a 500-km orbit, which primary mission was quantum-communication experiments, but a lasercom experiment called MCLCD (Coherent Laser Communication Demonstration) was planned as well [44]. The MCLCD space terminal shared the main optics with the quantum experiment used to transmit a 2.2-W 40-µrad 1549.731-nm laser beam with DPSK achieving a data rate of 5.12 Gbit/s in the 1.2-m Cassegrain telescope on the ground.

8.3.2 Applications

The most important application of lasercom from LEO is direct-to-ground downlinks since the main point is being able to download to Earth the increasing amount of data that remote-sensing missions require.

The resolution of these sensors is continuously improving, thus demanding more and more bandwidth. Moreover, as launches to LEO get more available, spacecraft get miniaturized, as is the case of CubeSats and other small satellites, and with several plans for constellations, the amount of data that will be required to be transferred from LEO to Earth is expected to grow dramatically in the coming years. The already-crowded RF spectrum will certainly not be able to support the growing demand for these communications requirements. Therefore, low-complexity and low SWAP (Size, Weight and Power) high-speed lasercom systems will be a key application to enable the use of such a big number of satellites in LEO.

A typical LEO lasercom link would be as follows: before the scheduled pass, an RF link is used to communicate with the satellite from the ground, updating the orbital data and other relevant pass information, and when the satellite is within the line of sight, it makes the tracking sensor face towards the ground station while a powerful beacon is transmitted from the OGS with a divergence wide enough to cover the uncertainty cone of the satellite position. This beacon is used by the satellite as a reference to close the tracking loop with the attitude control of the satellite (body-pointing), with the coarse-pointing system (typically a gimbal) and/or fine-pointing system (in case there is one, usually based on fast-steering mirrors) until the beacon is lost or the communication gets degraded.

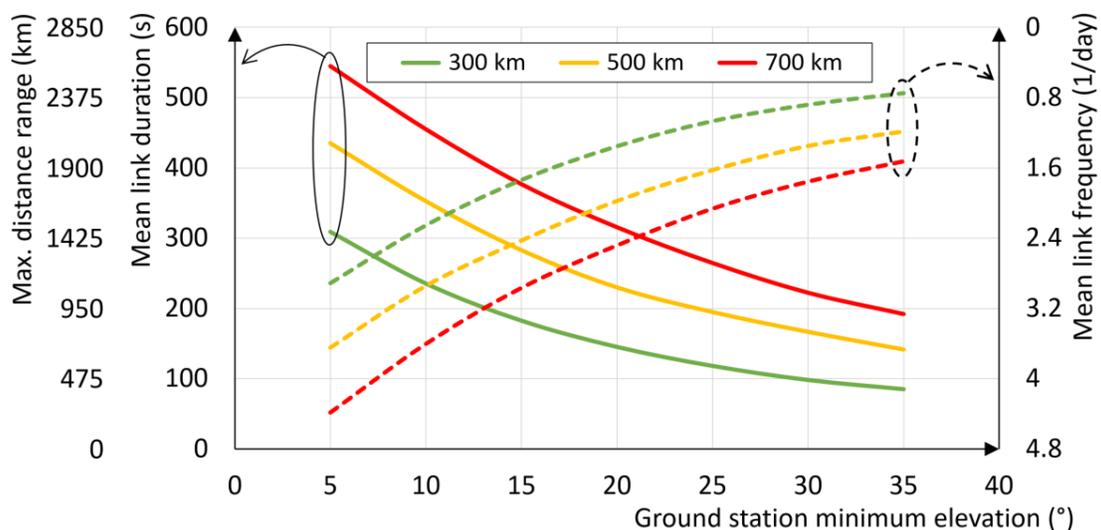

Fig. 8.20. Dependence of the mean link duration and frequency and the maximum distance range with the ground station elevation for three LEO altitudes at NICT's OGS at Koganei (Tokyo, Japan).

The LEO scenario generally implies frequent but short passes over a given ground station. Frequency and duration strongly depend on the maximum link range for a given orbit altitude, which is related to the minimum ground-station elevation. The Fig. 8.20 shows the dependence of the average link duration and frequency, and the maximum link distance with the ground-station elevation for three typical LEO altitudes. Although these parameters vary with the ground-station latitude, in this case the NICT's OGS at Koganei (Tokyo, Japan) was used, which is in an average latitude (35°41'58'') in the Northern Hemisphere. Taking the 500-km orbit as a typical example of LEO (Fig. 8.20, yellow lines), the maximum contact time would be ~7 minutes for a 5° minimum elevation, which implies a ~2,000 km maximum distance.

Another important limitation to lasercom through the atmosphere is the presence of clouds. Most types of clouds make optical links impossible in practice. Since LEO links are very short, the overall link availability is seriously limited when there are clouds at the time of the experiment. One solution to increase the availability of LEO missions is to relay the data through a GEO satellite. This important application of GEO lasercom can be found in the section 8.4. When access to a GEO relay is not available or data must be transferred directly to ground (due to a required short delay or because the onboard technology cannot close a link with GEO), site diversity is usually suggested as the key solution to maximize the probability of getting cloud-free lines of sight. However, for site diversity to work effectively, two conditions must be met: on one hand, there must be a network of decorrelated ground stations throughout the globe, and on the other hand, all of them must use a common set of specifications regarding the wavelength, modulation, codification, etc. In practice, currently neither of these conditions are generally true. Although there is an increasing number of OGSs in many different locations, they are typically built to support specific missions, usually demonstrations, and cannot easily interoperate with each other. The Optical-Communications Working Group of the Consultative Committee for Space Data Systems (CCSDS) is currently making efforts to produce standards to enable cross support between different agencies. A big and increasing number of LEO missions could benefit from being able to interoperate with many OGSs spread throughout the world, regardless the nature of their owners, be space agencies, universities or private companies. Furthermore, such cross-support would enable many more low-cost LEO missions which could not afford a dedicated OGS.

8.3.3 Space segment

A typical LEO lasercom terminal basically consists of an optical head assembly and an electronics/processing assembly. As shown in Fig. 8.21, it typically includes a telescope to transmit a collimated beam towards the ground and receive the uplink beam, some device to separate downlink and uplink (typically, based on wavelength, polarization, or both), a laser source (with or without amplification), an OOK external modulator (if data rates over Gbit/s are required; otherwise, the laser diode can be directly modulated by the input current in a simpler configuration), a tracking detector for the uplink beacon (based on 4-quadrant detector or on focal-plane array), a fast-steering mirror for the fine-pointing and another for the point-ahead angle if necessary, a coarse-pointing assembly (a 2-axis gimbal, or alternatively, the satellite attitude control), and optionally a fast photodetector for the communications uplink, if needed. The processing unit typically adds some codification against errors and interleaving against fadings, and controls communication, telemetry, command, and pointing system.

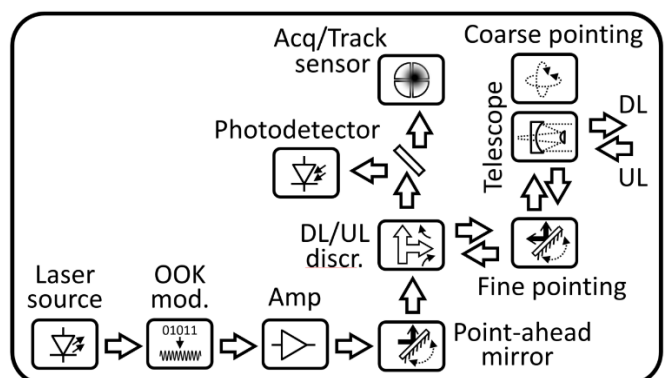

Fig. 8.21. Basic diagram of the optical-head assembly in a typical LEO lasercom terminal.

Lasercom space terminals usually implement one of the following three kinds of telescopes: In very-small terminals, e.g. mounted on CubeSats, based on downlink only, a simple lens (Fig. 8.22, left) can produce a collimated beam with a divergence narrow enough for the pointing system, especially when no fine pointing is implemented. For example, a near-diffraction-limited lens with a diameter in the order of 1-2 cm could suffice for a body-pointing-based small satellite. When bigger apertures are required, whether it is to produce a very narrow beam supported by a fine-pointing system, or to enable a communications uplink, reflective systems are generally selected. The classic Cassegrain configuration (Fig. 8.22, center) is a very common solution to build small and light assemblies with narrow field of view and good off-axis light rejection. When the receiving aperture must be maximized for the communications uplink, other good solution is an off-axis configuration (Fig. 8.22, right), which allows to make an effective use of the whole aperture, avoiding the central obscuration of on-axis reflective telescopes, and provides excellent off-axis light rejection.

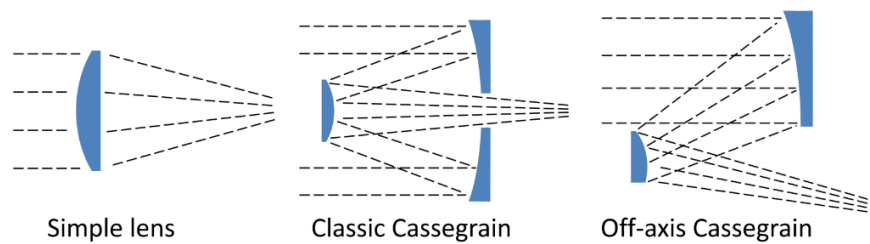

Fig. 8.22. Typical configurations for the telescope of lasercom space terminals.

The main challenge for LEO lasercom, with short passes and fast satellite motion, lies in the pointing and tracking system. In terms of beam divergence, equation (8.2) shows that a relatively-small transmitter's aperture can produce a very collimated beam that translates into a small footprint reaching the receiver on the ground. For example, a 5-cm aperture would produce a footprint in the order of 20 m from a distance of 500 km, or in the order of 100 m with an aperture as small as 1 cm. Considering a typical fine-pointing accuracy of 10 μrad and assuming a diffraction-limited transmitting telescope, according to the equation (8.18), the 5-cm aperture would introduce a pointing loss of less than 1 dB, or almost negligible in the case of the 1-cm aperture. With state-of-the-art technology, the pointing accuracy can go down to 1 μrad, allowing a very-low pointing loss even with very-narrow beam divergence.

A spectral filter of several nm before the fine-tracking detector can enable the lasercom terminal for daylight operation, while adding a low-rate modulation to the beacon can improve the background rejection as well as suppress the downlink scatter.

Acquisition and tracking detectors are typically based on 4-quadrant detectors (4QD) or on focal-plane arrays (FPA). The former detectors are based on 4 active photodiodes used to estimate the spot's center of gravity as the difference in the amplitude of the 4 elements, and the latter are camera-like sensors based on an array of pixels to directly image the beam. As a rule of thumb, FPAs provide a high spatial resolution, while 4QDs provide a high bandwidth. Additionally, 4QDs allow using modulated beacon, and the impact of radiation is lower. For their simplicity, 4QDs have been widely used as tracking detectors for a long time. However, the latest improvements of FPA technology have been significant, and it might become a good alternative to 4QDs, with state-of-the-art achieving extremely-high sensitivities. Although special care must be taken to protect FPAs against radiation because of their higher vulnerability compared to 4QDs, their applicability would be more favorable in LEO, where the radiation environment is not as severe as in other scenarios farther from the Earth's magnetic protection.

As explained in 8.1.2, depending on the beam divergence, it may be necessary to transmit the downlink beam with an offset angle with respect to the detected uplink beam due to the relative motion between the satellite and the Earth, and the finite speed of light. As shown in the equation (8.4), this point-ahead angle is variable depending on the tangential velocity of the satellite as seen from the ground station. For LEO-to-ground links, the maximum velocity, thus the largest point-ahead angle, will happen when the satellite is at the zenith. Taking a worst-case example to illustrate this, a 500-km LEO satellite at the zenith will travel at a velocity of 7.62 km/s, which requires a point-ahead angle of 51 µrad.

Fig. 8.23 shows three possible configurations for the point-ahead mechanism. In the left-hand side diagram, the uplink position is detected by a 4QD after separating uplink and downlink, and the downlink pointing is monitored by another 4QD after the point-ahead mirror (PAM), which makes it possible to compute the required point-ahead angle in real time. In the diagram in the center, a small fraction of the transmitted downlink is sent back to the 4QD by using a Corner Retroreflector (CRR) when there is no downlink signal in order to calibrate the necessary point-ahead angle. Real-time point-ahead angle calculation could be implemented if an FPA is used instead of the 4QD and the power is carefully controlled to produce similar received intensities in the FPA. The right-hand side diagram is based on a pre-calibration without monitoring the transmitted downlink by using the beacon from the ground, which despite its simplicity may be a difficult process with fast-moving LEO satellites. When real-time point-ahead angle calculation is not possible, a reference-frame transformation is required to obtain the precise direction the point-ahead angle must be applied towards.

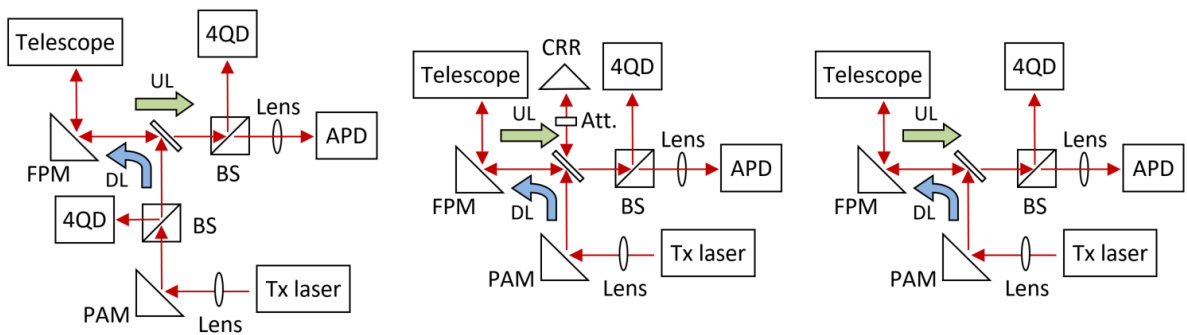

Fig. 8.23. Three possible configurations for the point-ahead mechanism in the space terminal.

In the LEO-to-ground scenario, the preferred wavelengths are around 1550 nm, in the C-band, because of the abundance of devices to generate the laser source, amplify and modulate it. Besides the commercial availability, the reduced atmospheric attenuation, sky-background level as well as turbulence effects make this wavelength the most common in this scenario. The wider beam divergence for the same transmitting aperture is not as important as in other scenarios because of the shorter distance. At this wavelength, the basic diagram to generate the downlink signal is shown in the lower-left side of Fig. 8.21. For a high-speed configuration, i.e. multi-Gbit/s, the most common scheme would consist of a low-power laser source, followed by an external modulator and an amplifier. The oscillator would be typically based on a semiconductor laser diode. This technology has matured in a significant way to produce high power and high efficiency. However, there is a tradeoff between transmitted power, beam quality, and modulation rate. Only if power and speed requirements are not demanding, a single laser diode can be used, directly modulated and with no amplification. Laser diodes producing several hundreds of mW are difficult to modulate at high speed (e.g. 1 Gbit/s or faster) and the output beam quality is insufficient for the requirements of external modulators.

If data rates in the order of 100 Mbit/s and transmitted powers in the order of several hundred mW are sufficient, directly modulated laser diodes can be a simple solution. When both high-speed multi-Gbit/s and high power above 1 W are required, the configuration shown in Fig. 8.21. is the most common one. Despite the additional insertion losses, external modulators are packed in compact devices and allow bigger extinction ratios and speeds over tens of Gbit/s. Output powers in the order of several W can be achieved when using fiber amplifiers, usually based on Erbium-doped fibers at 1550 nm, i.e. EDFA (Erbium-Doped Fiber Amplifier). When high output power is required, several EDFA amplification stages can be used, separated by optical isolators. Since fiber amplifiers typically use single-mode fibers, the output beam quality is excellent, based on a single spatial mode, and the alignment is stable and easily controlled with the fiber connector. The wall-plug efficiency (ratio of the total optical output power to the input electrical power) in the order of 10% is the main weak point of EDFA technology, being one of the most power-consuming components in the lasercom system. When available power in the satellite is a strong constraint, using a wavelength around 1 µm could be a solution, since the wall-plug efficiency of YDFA (based on Ytterbium) is more than 2 times better than EDFAs.

8.3.4 Ground segment

The optical ground station has two important roles in a LEO lasercom system: on one hand, it must provide an uplink beacon so that the satellite can track the ground telescope to transmit its downlink signal, and on the other hand, the OGS must be capable of receiving the communication downlink beam from the satellite and demodulate/decode the received signal.

The first role is simpler to implement than the second one. Generally, the beam divergence of the beacon laser should be wide enough to cover the uncertainty cone of the satellite position at its maximum distance, determined by the minimum OGS elevation. For example, a typical uncertainty error in the order of 1 km would require a full-angle divergence over 1 mrad for a LEO satellite at a distance of 1000 km. The optics the beacon is transmitted through are usually mounted in parallel to the receiver's telescope and aligned with its tracking detector. Due to its wide divergence, the aperture is usually small (in the order of a few cm), and the transmitted power is usually rather high (typically between several W and tens of W). Because of the high transmitted power, an eye-safe wavelength is preferred, e.g. in the lower part of 1.5-µm C-band or in L-band, leaving the upper part of C-band for communications, where EDFAs are more efficient and produce lower noise figures, which is more important for communications than for beacon.

A common strategy to reduce the received-power scintillation in the satellite is to transmit several parallel beams [45]. If they are separated by a distance longer than the atmospheric coherence length, the effect of the atmospheric turbulence is averaged in the combined signal, and the variance of the power fluctuation reaching the satellite is reduced. This technique is more common for the beacon than for communication uplinks, since the beacon is transmitted with a wider beam divergence to cope with the satellite's position uncertainty and the communication beam is narrower allowing a more effective power coupling into the receiver. In case of beacon, the intensity of the transmitted optical power can also be modulated for a better background-noise rejection in the satellite's tracking detector. By using a bandpass filter around the carrier frequency (typically around several kHz or tens of kHz), all the low-frequency and DC components present in the time-varying background noise can be eliminated.

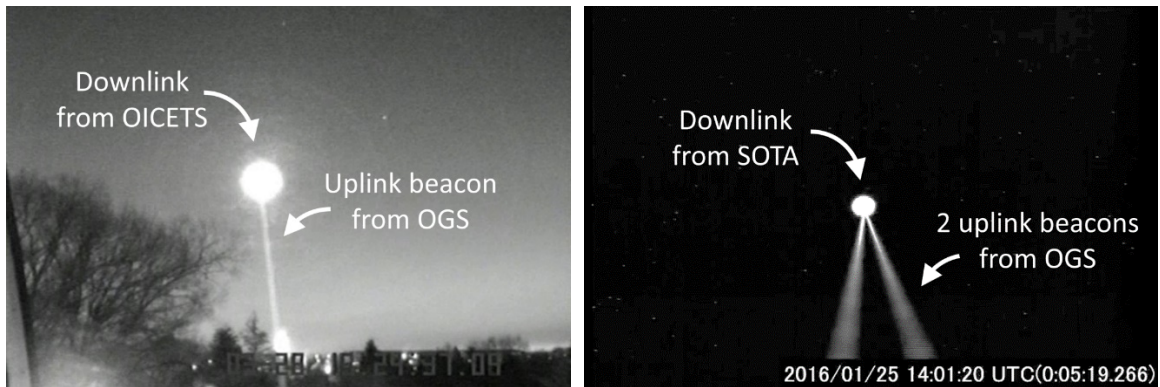

Fig. 8.24. Examples of uplink-beacon transmission from 2 NICT's LEO-to-ground missions: 808-nm beacon towards OICETS (left) and 1064-nm beacon towards SOTA (right).

Fig. 8.24 shows images taken with IR cameras of the uplink-beacon transmission from two NICT's LEO-to-ground missions. On the left, the beacon towards OICETS can be seen as a straight line in an experiment performed on 28 March 2006, as well as the downlink from the satellite as a bright dot. On the right, the beacon laser towards SOTA is shown, as well as the downlink signal from the satellite in an experiment performed on 25 January 2016.

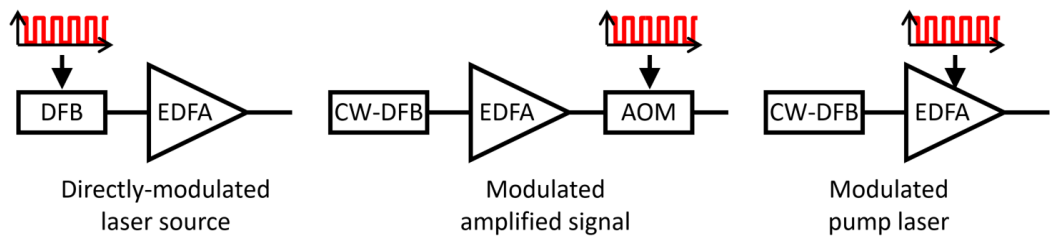

Fig. 8.25. Typical modulated beacon laser configurations.

Regarding the beacon technological implementation, since the transmitted power is high, the usually preferred wavelength is 1550 nm due to its eye safeness. At this wavelength, the typical beacon laser configuration would be based on one of the designs shown in Fig. 8.25, depending on how the modulation is introduced in the beacon signal. The design on the left is the simplest one, based on modulating the seed laser directly by controlling its input current. It is applicable with modulation frequencies of several tens of kHz. For lower frequencies, the excited Erbium ions start to lase with no seed signal causing parasitic oscillation which may lead to a permanent damage. However, there are partial solutions to this even at those lower frequencies, namely, if a dummy CW signal is introduced at other wavelength (although the total gain would be reduced) or if the modulation depth is lower to 100% (which makes the demodulation more difficult in the receiver). When a lower modulation frequency is required, the design on the center is a good solution, based on modulating the amplified signal with an Acousto-Optic Modulator (AOM), which allows a wide range of modulation frequencies, at the cost of a lower output optical power, since the AOM sets a maximum input power in the order of 3W. The design on the right consists in modulating the pump laser of the EDFA, and offers high output power and low modulation frequencies, typically below 5 kHz.

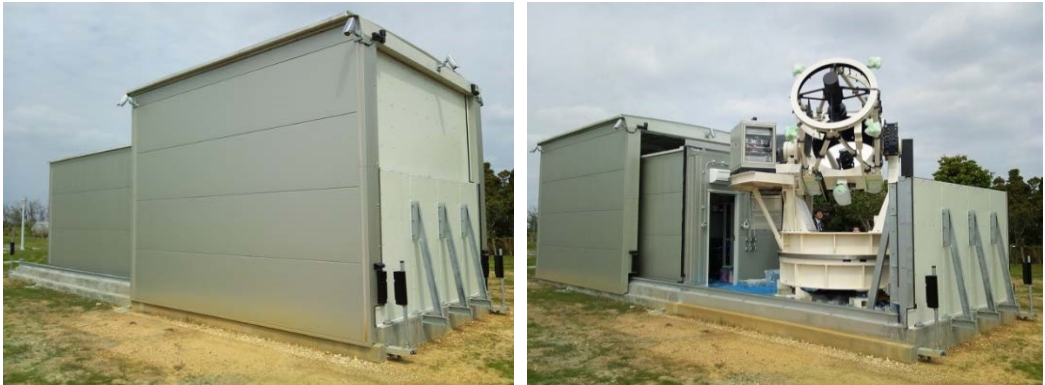

Fig. 8.26. Compact and remotely-controlled 1-m OGS in Okinawa (Japan) built by NICT.

The second function of the ground segment in a LEO mission is receiving the communication downlink from the satellite. In LEO-to-ground links, the aperture size of the receiving telescope is an important design parameter. On one hand, the link budget benefits from the bigger gain that large apertures provide, but on the other hand, the cost of building monolithic mirrors scales as aperture to the 2.5 power [46], which makes them expensive. In addition, big telescopes incur on other costs such as operational costs and a dedicated facility. Furthermore, a pointing system that allows a big telescope track LEO satellites is complex and expensive. For all these reasons, considering the latest improvements in the sensitivity of the receivers, together with the fact that the link budget in LEO-ground links is not as demanding as in other scenarios, the tendency has gone towards reducing the aperture size of the ground telescopes. Fig. 8.26 shows an example of a 1-m compact OGS built by NICT in Okinawa (Japan), which can be controlled remotely from NICT headquarters in Tokyo.

In typical LEO-to-ground lasercom links, where the link budget is not as constrained as in other scenarios, OOK modulation is often used. Even though DPSK offers a 3-dB advantage over OOK, the latter is usually preferred because of the simplicity of the receiver. Whereas a simple photodetector is enough to demodulate the OOK signal, DPSK requires the use of a balanced detector with an interferometer adapted to the specific data rate (as it was explained in section 8.1.2). Furthermore, in LEO-to-ground links, the fast motion of the satellite during the pass produces a Doppler shift in the order of several GHz that can degrade the demodulation process if the path-length difference in the interferometer is not precisely controlled by tracking the received wavelength, adding complexity to the system. In addition, the simplicity of OOK receivers allows using a multi-mode fiber to couple the received signal into, which is relatively insensitive to atmospheric turbulence, as opposed to single-mode fibers due to their small core diameter, i.e. smaller than 10 μm in single-mode fibers and bigger than 50 μm in multi-mode fibers. Moreover, if the area to couple the signal into is bigger, a bigger PSF can be tolerated, which enables the use of low-cost telescopes with reduced optical quality.

Despite the coupling technique the receiver uses, some scintillation effect will usually remain, since this depends mainly on the conditions of the atmospheric turbulence, as explained in the section 8.2.2. Together with fadings due to pointing errors from the space terminal, the communication link might require the use of FEC coding, depending on the received power and the sensitivity of the receiver. The quantum-limited sensitivity of OOK is 33.9 photons/bit [47]. However, thermal noise makes practical implementations show worse sensitivities. As shown in Table 8.3, typical minimum required sensitivities for uncooled-APD-based receivers at 1550 nm are in the order of 500 photons/bit at 1 Gbit/s to achieve a BER better than $10^{-3}$ (which can be further improved with FEC), or 1000 photons/bit to achieve a BER better than $10^{-9}$ (with no FEC). This kind of receivers allow multi-mode fiber coupling for the received signal, which is a simple technique, usually only requiring tip/tilt correction with a fast-steering mirror to compensate the beam wander caused by atmospheric turbulence (low Zernike modes) and improve pointing stability (as explained in section 8.2.2).

Table 8.3. Typical sensitivities to achieve a certain hard-decision BER for different receiver's schemes at 1 Gbit/s and 1550 nm.

| Receiver's scheme | BER (w/o FEC) | Minimum required power |
| --- | --- | --- |
| Uncooled-APD | $10^{-9}$ | < 1000 photons/bit (-38.92 dBm) |
| Uncooled-APD | $10^{-3}$ | < 500 photons/bit (-41.93 dBm) |
| Optically-preamplified | $10^{-9}$ | < 100 photons/bit (-48.9 dBm) |

More complex receiver's schemes based on optically pre-amplification, i.e. low-noise EDFA before the photo detection, can take the sensitivity from 1000 photons/bit in uncooled APDs down to about 100. However, single-mode fiber coupling is required in order to use EDFAs, which makes the receiver more complex, generally requiring adaptive optics to correct the aberrated received waveform. For this reason, this scheme is not usually considered for LEO-to-ground links, where the link budget is not as constrained as GEO-to-ground, where it finds its natural application, as explained in section 8.4.4. The APD scheme is applicable up to several Gbit/s, but to achieve higher data rates the optically preamplified scheme should be considered, or alternatively and probably simpler, Wavelength-Division Multiplexing (WDM).

8.4 Geostationary satellite communications

Satellites at Geostationary Equatorial Orbit (GEO) appear immobile for observers on ground, because the satellite rotates synchronously with the Earth. This property makes GEO satellites especially suitable for communications, streaming or weather monitoring. GEO satellites are interesting because of their large coverage: three satellites can provide world-wide coverage [48]. The classical GEO-satellite communications market is focused on video broadcasting and internet access. In recent times, this has been enhanced by backhaul service for edge servers buffering most popular videos, and direct user access to broadband Satcom services via VSAT terminals and High-Throughput Satellites (HTS). HTS can offer communications throughput of more than 100 Gbit/s. Current technologies are based on Ka-band and Ku-band technology. Optical communications in GEO have been mainly developed for data-relay from LEO, and its major example is the ESA EDRS system. For Very High-Throughput Satellites (VHTS), big amount of data needs to be transmitted to the satellite. New communications satellites may require data throughputs in the order of Tbit/s and more for the feeder-uplink; i.e. the link between the ground station and the satellite. Optical feeder links may become the next revolution in space, boosting the available data throughput with a potential global coverage using few satellites.

8.4.1 Heritage

ETS-VI: the first GEO-to-ground lasercom demonstration

The Japanese Engineering Test Satellite VI (ETS-VI) (shown in left image of Fig. 8.27) included the first lasercom terminal in GEO orbit called LCE (Laser Communication Equipment). It was developed by the NICT's Communications Research Laboratory, currently NICT's Space Communications Laboratory and launched on 28 August 1994 by the National Space Development Agency of Japan (NASDA, currently JAXA). The objectives of this mission were to evaluate for the first time the key technologies for satellite optical communications. This downlink signal could transmit an onboard pseudo-random noise sequence to the ground station, relay the uplink signal back to the ground station, or transmit the telemetry data from different lasercom components on LCE at 128 kbit/s at an 8× redundancy to achieve the 1.024 Mbit/s [2] [49]. The LCE lasercom terminal onboard ETS-VI is shown in right image of Fig. 8.27.

In 1995, the ETS-VI satellite was used to carry out a joint experiment with NASA-JPL, which used 2 separated OGSs in Table Mountain (Wrightwood, California) to communicate with the satellite during the project GOLD (Ground/Orbiter Lasercomm Demonstration) [50].

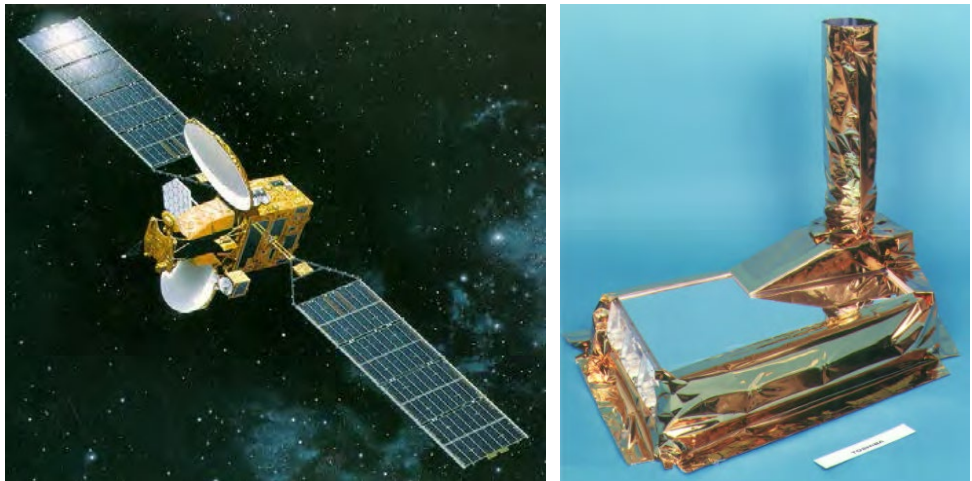

Fig. 8.27. ETS-VI satellite (left). LCE lasercom terminal onboard ETS-VI (right).

After the first demonstration with ETS-VI, two missions with an optical terminal onboard were launched to GEO: the Artemis satellite in 1998 by ESA, which will be described in the next section, and the GeoLITE satellite in 2001 by the National Reconnaissance Office (NRO) [51], although there is no published information about this mission after its launch, meaning that probably it did not succeed.

Artemis satellite and the first inter-satellite experiments

Semiconductor laser inter-satellite link experiment (SILEX) was the first civilian optical communications program for space (in the frame of ESA DRTM) [3] [52] and the first step towards the European Data Relay System (EDRS). Inter-satellite links were the framework of the SILEX project with the main objective of relaying video data from a LEO satellite to a ground station, demonstrating the feasibility and performance of optical inter-satellite links in space [3]. The experiments involved two satellites which hosted the optical terminals: the ARTEMIS GEO satellite and the SPOT-4 LEO satellite. SPOT4 was an Earth-observation satellite, developed by Matra Marconi Space for CNES. It was successfully launched in 1998 and ARTEMIS, developed by Alenia for ESA, in 2001. The first image transmission was carried out in 2001 between the two optical terminals (see Fig. 8.28, left), from PASTEL on SPOT-4 to OPALE on Artemis and then to the SPOTIMAGE ground station in Toulouse via Ka-band feeder link [53].

Table 8.4. Main parameters of the optical terminals on ARTEMIS and SPOT-4 satellites.

|  | ARTEMIS | SPOT-4 |
|---|---|---|
| Antenna diameter (mm) | 250 | 250 |

| Beam diameter TX (1/e$^2$) (mm) | 125 | 250 |
|---|---|---|
| Transmit power (mW) | 5 | 40 |
| Transmit data rate (Mbit/s) | 2 | 50 |
| Transmit wavelength (nm) | 819 | 847 |
| Transmit modulation scheme | 2-PPM | NRZ |
| Receive data rate (Mbit/s) | 50 | - |
| Receive wavelength (nm) | 847 | 819 |
| Receive modulation scheme | NRZ | - |
| Beacon wavelength (nm) | 801 | - |
| Optical terminal weight (kg) | 160 | 150 |

The laser terminals were developed based on OOK modulation and direct detection of laser beams in the 800-nm range (GaAlAs laser diodes and an APD-based receiver). The main parameters of both terminals involved in SILEX experiments are summarized in Table 8.4. The terminals allowed 50-Mbit/s data-rate transmission. The terminals on both satellites (OPALE and PASTEL) had a similar structure: a fixed-part electronics and a satellite interface structure with a mobile part. The electronics comprised the onboard processor, the coarse-pointing drive electronics and the communications electronics (interfacing with the signals coming/going from/to the LEO/GEO terminal. The satellite interface structure carried the coarse-pointing mechanism that moved the mobile part, formed by the telescope, the focal plane (with sensitive elements, as sensors and sources) and the required electronics. Due to the relative motion between the satellites, a point-ahead angle assembly with a piezo-electrical mirror for the fine tracking was included in the optical system. A high-power laser beacon was used during the acquisition phase for the partner detection on the GEO terminal. The beacon scanned around 750 μrad in the direction of PASTEL. When it was illuminated by the beacon, PASTEL corrected its angle pointing the communication beam to OPALE which used the incoming wave to close the loop [3] [52] - [54].

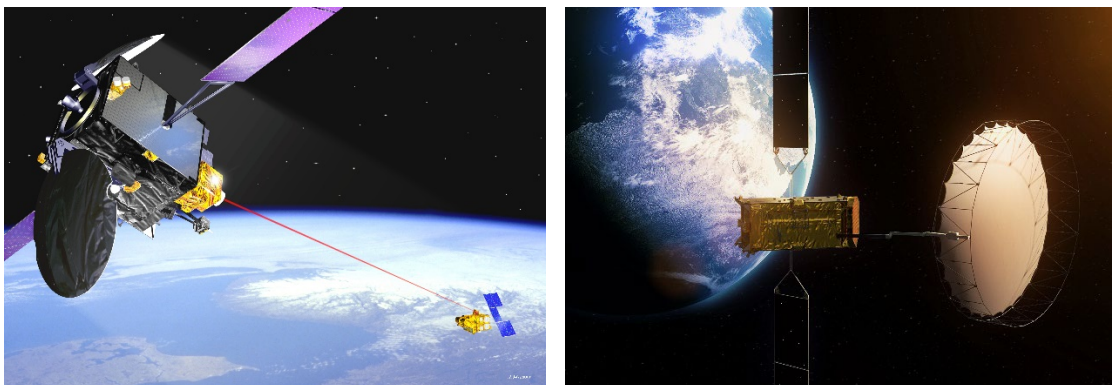

Fig. 8.28. Representation of ARTEMIS and SPOT-4 satellites (left) and representation of Alphasat (right).

Moreover, since November 2001 bidirectional links were established between ARTEMIS and the OGS on Canary Islands. During these experiments BER and atmospheric parameters were measured [55].

Other inter-satellite links were performed between ARTEMIS and the Japanese satellite OICETS. OICETS was a relatively-small satellite with a mass of approximately 570 kg. The lasercom terminal on the satellite was designed to have visibility of the geostationary satellite ARTEMIS. The optical antenna had a primary mirror with an effective diameter of 26 cm in a center-feed Cassegrain configuration and the power consumption of the terminal was around 320 W [56] [57].

Inter-satellite laser communications between OICETS and ARTEMIS were performed several times by JAXA and ESA since 2005, when the first bidirectional inter-satellite link took place [57].Inter-satellite links were also performed through the atmosphere, with the link maintained until the Earth surface blocked the communication. This way, the atmospheric influence on the transmitted beam was measured and the beam pointing and tracking errors were analyzed as well. [58].

Alphasat, the operational inter-satellite links and the news on relay systems

In July 2013, the Alphasat satellite was launched and located at the GEO orbit 8°E. This satellite carries several demonstration payloads for satellite communications, and among them an optical terminal developed by TESAT Spacecom for LEO-GEO inter-satellite links (see Fig. 8.28, right). The lasercom terminal (see Fig. 8.29) is a made of a central rectangular base structure with a coarse pointer (gimbal) and the optics unit. The telescope aperture is 135 mm [59] and the tracking and the communications receiver is based on homodyne reception at 1064 nm. The main parameters of the optical terminal are summarized in Table 8.5.

Table 8.5. Alphasat optical payload main parameters.

| | |
|---|---|
| Dimensions (cm) | 60×60×60 |
| Weight (kg) | 50 |
| Power consumption (W) | 160 |
| User data rate (Gbit/s) | 1.8 |
| Operation range (km) | 38,600 |
| Transmit power (W) | 5 |
| Wavelength (nm) | 1064 |
| Transmit Beam diameter (mm) | 135 |
| Transmit Beam divergence (μrad) | 10 |
| Telescope Diameter (mm) | 135 |
| Nominal receive power density (μW/m$^2$) | 280 @ BER 10$^{-8}$ (1.8 Gbit/s) |

The EDRS design is originally constituted of four satellites: two over Europe (EDRS-A and EDRS-C), one over America (EDRS-B) and one over Asia-Pacific (EDRS-D). In 2018, EDRS-A and EDRS-C are fully deployed, providing operational links since November 2016 for the Sentinel Earth-Observation LEO satellites under the "Space Data Highway" service operated by Airbus [60]. The EDRS-B satellite over America will be most-probably not developed. Instead, the American LCRD system will be deployed. The EDRS-D over Pacific is at this time planned and it will provide interoperability with the 1550 nm wavelength, 3.6 Gbit/s data transmission, GEO-GEO cross-link communication with the others EDRS satellites and it will allow bidirectional links with aircrafts. The goal is to create a network that interconnects GEO satellites to LEO and aircrafts (or pseudo-satellites) optically and transfers the information from GEO with radiofrequency to the ground. In addition, there are studies to extend the use for data-repatriation from ground [61]. This last scenario is especially interesting for isolated areas, where no ground infrastructure is available or it is limited. An OGS can send back to Europe (in this case) the data over the GEO satellite, avoiding long time delays, like for example in the case of Antarctica [62].

The EDRS is being proposed and it is currently being discussed for a possible standardization in the CCSDS Optical Communications Working Group. In this organization, the major international players in space including the space agencies and the industry works mainly in three scenarios: High Date Rate, Low-Photon Flux (Deep-Space Communications) and Low Complexity (direct LEO downlinks) [63].

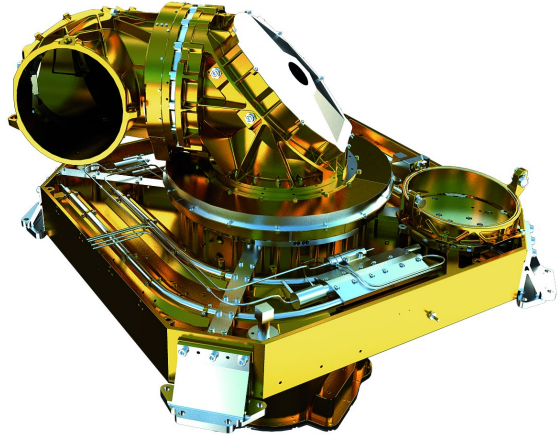

Fig. 8.29. Operation GEO optical communications terminal from Tesat Spacecom.

The forthcoming systems

Currently, NASA is developing its own relay system called LCRD (Laser Communications Relay Demonstration), which should start operating in 2019 [64] [65]. In Japan, JAXA is developing another relay system called JDRS (Japan Data Relay System) to start operating in 2019 [66] [67], and NICT is developing a GEO feeder-link terminal called HICALI (High-speed Communication with Advanced Laser Instrument), which aims at demonstrating bidirectional lasercom from GEO up to 10 Gbit/s in 2021 [68] [69]. These communications systems are also being standardized in the Optical Working Group of CCSDS under the High Data Rate scenario.

LCRD is the first step towards an American relay system for supporting human exploration and advanced instruments aboard science missions. It is a joint project between NASA's Goddard Space Flight Center, the Jet Propulsion Laboratory, the California Institute of Technology and the Massachusetts Institute of Technology Lincoln Laboratory. The space segment is constituted by two optical terminals onboard the spacecraft and one high-capacity radiofrequency terminal will relay data from and to other satellites, spacecraft, airplanes and ground stations. The ground segment includes three OGSs, two optical in Hawaii and California and one in Ka-band in New Mexico. Two Mission Operations Center will be connected to the ground stations by terrestrial links. The ground stations will be used to simulate spacecraft users with specific daily data volume requirements. The optical links will provide a bidirectional 1.244 Gbit/s data rate. The Ka band will support one or two users at 32 Mbit/s in forward-link and one user at 622 Mbit/s or two users at 311 Mbit/s in return link. Each optical terminal is constituted by a telescope, the electronics for pointing and acquisition, and a modem that supports PPM and DPSK in both link directions. The experiment will also demonstrate especially-developed encryption technology for Information Assurance. Modems supporting both modulation formats were developed and demonstrated in the past years, under the LADEE mission, which demonstrated in 2014 optical PPM data communication from the Moon, achieving up to 622 Mbit/s of user-data rate.

In 2019, JAXA plans to launch the first satellite of the Japanese Data Relay System (JDRS). JDRS will include a feeder link in Ka band and in optical it will provide data rates of 2.5 Gbit/s for return link and 60 Mbit/s for forward link including FEC codification, leading to a user data rate of 1.8 Gbit/s and 50 Mbit/s respectively. The inter-satellite communication system is designed in the optical C-Band, 1540 nm for the forward link and 1560 nm for the return link. The optical terminal in GEO will have an antenna diameter of 15 cm, whereas in LEO the aperture will be 10 cm. The acquisition is based on a beaconless sequence to be completed within 60 seconds. The optical communication system is based on direct detection, with DPSK for the return link and intensity modulation for the forward link. The first JDRS user will be the Advanced Optical Satellite, a Japanese optical observation LEO satellite from JAXA.

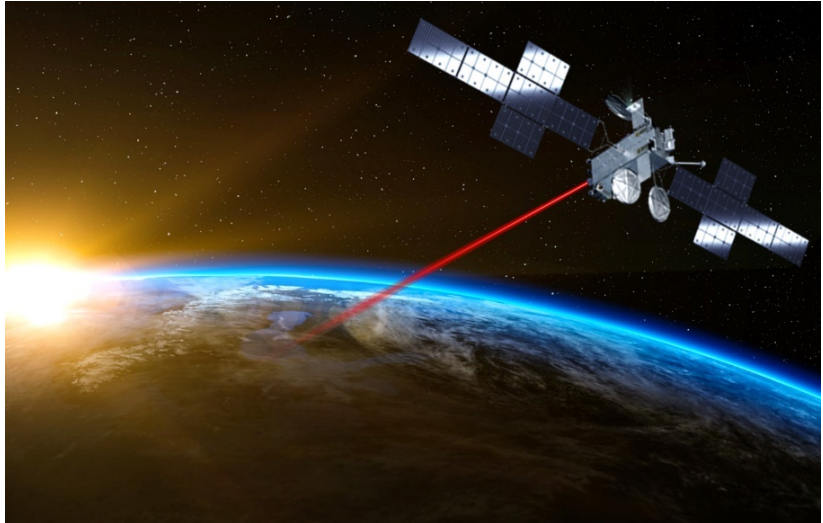

Fig. 8.30. Artistic illustration of NICT's HICALI onboard ETS-IX.

NICT in Japan is developing a high-throughput GEO satellite called ETS-IX based on hybrid use of radio and optical frequencies (Fig. 8.30). ETS-IX will be launched in 2021 in the second flight of the new Japanese launch vehicle H3. The Ka-band system will include feeder links and user links with 100 Mbit/s per user with flexible allocation of frequencies and steerable beams to handle traffic fluctuations. The lasercom terminal is called HICALI (HIgh speed Communication with Advanced Laser Instrument) and it will demonstrate a bidirectional 10-Gbit/s feeder link between GEO and ground. The HICALI terminal will transmit a 2.5-W 1540-nm laser through a 15-cm aperture to be received on the ground by a 40-cm receiver's aperture. For the uplink, a beacon system consisting in 4 apertures of 5 cm will transmit a total power below 20 W at 1530 nm, and the feeder uplink will transmit 2.5 W at 1560 nm through an aperture smaller than 40 cm. The HICALI mission will also include feeder-link experiments by using the NICT's OGS network to demonstrate the high availability provided by site diversity, allowing fast handovers between different ground stations depending on the cloudiness over each site.

8.4.2 Applications

For optical communications, satellites at this orbit have been proposed mainly for data relay from LEO. Transmission to GEO satellites extends the ground coverage and decrease the delay between users which are not visible from the same satellite. Moreover, the intermittence of the LEO satellites visibility is overcome by transmitting to GEO satellite, reducing the on-board data recorders and increasing the capacity. On the left of Fig. 8.31 there is a representation of a satellite relay system. The coverage of the lower orbits is larger at the GEO than from a given station on the ground. For this application, the ESA developed the EDRS, which became operational on November 2016, giving service to the European Commission's Earth observation program.

Classical satellite communication is mainly focused on television broadcasting, but in the forthcoming years, providing internet access may gain more weight. The optical-fiber infrastructure is costly, especially in less-dense populated areas. Several initiatives in Europe target a full coverage of Internet access at 50 Mbit/s and more, which would not be reached only with the terrestrial infrastructure. For example, currently the 30% of the rural areas in Germany are not covered. In this sense, satellite communications can be a good complement to reach the full coverage to broadband internet access. Another aspect is 5G, the Industry 4.0 and the Internet-of-Things, which require also global connectivity. In this case, new business models based on cloud services and real-time monitoring of the production and transport of products will require internet access everywhere and at any time. Geostationary communications time delay of at least ~250 ms due to the signal propagation may be limiting for applications like telepresence or augmented reality, where a direct human interaction is expected, but for other kind of applications, it may not have any impact.

On the right of Fig. 8.31, a representation of the GEO satellite-based communications is shown. A bidirectional link, called feeder-link, between the satellite and a ground station, called gateway, connects the satellite to the network. The satellite gives connectivity to the surface with several spot beams, called user-links. The feeder-link therefore carries a lot of traffic, with throughputs that currently reached several hundreds of Gbit/s, but in the future it can potentially be of several Terabit/s [70]. Current technology based on radiofrequency is reaching to the limit on providing the required throughput due to the limited available bandwidth. Hence, several OGSs operating in the same spectrum feed the satellite to reach the required traffic. The number of these stations increases linearly with the throughput, reaching hundreds of them when approaching Terabit/s. Therefore, in the future, communication satellites will integrate optical communications for the feeder link, to increase the data throughput. This would have two advantages, first the feeder-link capacity will not be limiting anymore, due to the ~10 THz of non-regulated spectrum around the 1064 nm and 1550 nm and second, the spectrum currently used for the feeder link could be allocated for the user links, increasing the overall throughput by these two means.

In order to make an effective use of the available spectrum in the optical domain, wavelength division multiplexing (WDM) is required. Optical channels centered at different optical frequencies are multiplexed together, being able to reach an aggregate throughput of multiple terabit/s. In fiber communications, such technology is state-of-the-art for the optical C and L bands. The center frequency of each channel is defined in an ITU standard for the Dense WDM, with channels bandwidth of 100 GHz, 50 GHz, 25 GHz and 12.5 GHz [71]. The use of such technology in the atmospheric turbulent channel was demonstrated in 2016, transmitting 1.72 Tbit/s in 40 channels with 100 GHz bandwidth using OOK modulation [72], and in 2017 transmitting 13.16 Tbit/s in 51 channels with 50-GHz channels using 16 QAM and QPSK modulation [73].

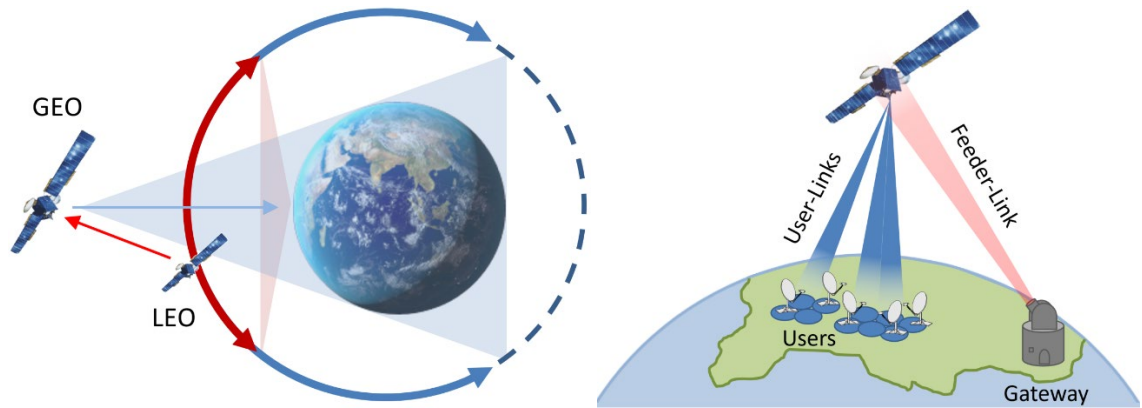

Fig. 8.31. Applications for GEO optical satellite communications. The GEO data relay provides a larger coverage to LEO for data repatriation (left). Satellite communications for internet access and streaming based on optical feeder-links (right).

Current projects on optical communications in space target one-purpose satellites. It means that only one application is addressed. In the future, if the demand on connectivity and bandwidth does not drop off, the ground network may be extended to the space. In this case, GEO satellites acting as nodes may connect directly to the ground network, as backhaul links. GEO is advantageous due to the relative-small movement of the satellite, compared to LEO satellites, which require handovers quite often. The GEO satellite nodes could then distribute the connectivity to other satellites in GEO, MEO or LEO for relay and telecommunications applications, being then fixed infrastructure in space as an extension of the terrestrial network. Designing the ground segment properly, continuous optical connection to the satellite could be guaranteed, meaning that most of current bottleneck due to the limited radiofrequency spectrum would be solved. This approach would be compatible with a LEO constellation, for example. Connecting this constellation with the GEO satellites through a relay link would extend the high data rates at all orbits and to any kind of satellites, even small LEO satellites like CubeSats [74], which could make use of a LEO constellation to relay the data through the GEO satellite. Traffic demands on small delays could be then routed anyway with a direct link to Earth.

8.4.3 Space segment

Communications systems on satellites are limited by the platform mass and power budgets and therefore its design is constrained to the available onboard resources. This makes the design of very-high data-rate systems for space very challenging. Currently, there are no satellite buses specifically designed for very-high throughput communications based on lasercom, specially targeting multiple Terabit/s. Therefore, there is still work to be done on the space segment to design appropriate platforms to accommodate such kind of terminals.

One important point to design future systems based on optical feeder links will be to keep the compatibility towards users, which requires including digital satellite television broadcast standard DVB-S2X for the user links. This standard defines the physical layer, including modulation and coding schemes. The user-link modulation is defined as APSK with different modulation orders, each one targeting different carrier-to-noise ratio requirements, being the lower orders for bad SNR conditions. For the current system technology in RF, it seems reasonable to use modulation orders up to 16 or 32 APSK, but the standard defines modulation orders up to 256 APSK.

From the communication system point-of-view, there are several approaches still under research for future-generation systems. The different options can be classified between analogue and digital payloads and between transparent and regenerative and they are summarized in Fig. 8.32. The choice is driven by a trade-off between complexity and efficiency.

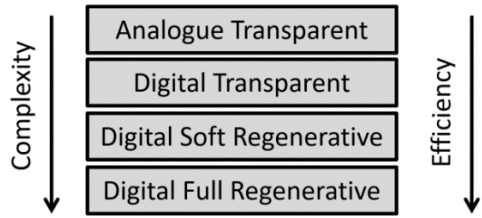

Fig. 8.32. Communications design approaches.

Currently, the satellite-communication community prefers analogue transparent systems, acting as transponders, as state-of-the-art satellite-communication systems based on radiofrequency are. This allows developing and upgrading the communication system without compatibility issues of previous satellites . From the point of view of the optical feeder link, that means that the data needs to be analogue modulated on the laser light [75], in a similar way as radio-over-fiber [76] to avoid any data processing onboard. APSK signals can be mapped in amplitude analogue modulation, by setting the signal into an intermediate frequency, which first would help to convert the signal on the satellite to the Ka band and allows multiplexing more DVB carriers in a DWDM channel. At the satellite terminal, the signal is converted to the electrical domain and frequency up-converted to Ka band for the user link. With this approach, no signal processing on the satellite is required, and therefore the ground segment must take care that the signals are shaped for the user link correctly.

Another approach that keeps the transparency of the system is to digitally sample the signal and transmit the samples on the feeder link. This approach is similar to the analogue one, but it allows adding FEC on the bits containing the sampling information for the feeder link to mitigate the turbulence effects in the uplink. This means that onboard processing is required for decoding the FEC. In this case, the system is transparent from the point of view of the DVB signal, since the wavefront is digitally sampled. The main drawback of this approach is the band expansion due to the sampling but in general it is more efficient than the analogue approach. Error correction algorithms help achieving a better sensitivity. An overview of this approach can be found in [77] [78].

Soft-regenerative techniques are on the half way between transparent systems and a full-regenerative system. In this case, the system is transparent to the data, but it requires a generation of the DVB wavefront for the user link. The constellation points of the DVB signal can be mapped to the optical feeder link signal. The band expansion depends on the modulation order of the optical signal, but it is lower than in the digital transparent option. Data is transmitted in baseband and error-correction algorithms can be used to protect the transmitted information. At the satellite, data should demap the constellation points of the feeder link into the APSK DVB format and the wavefront must be generated. However, there is no need to access to the DVB frame.

Finally, a fully regenerative system would be the most efficient approach, allowing optimizing the physical layer. This option requires a full manipulation of the contents of the DVB signal and it would require completely generating the signal onboard the satellite. That means that this is the most-costly option in terms of signal processing and therefore on mass, power consumption and heat dissipation.

However, although this option would exploit the full potential of the satellite link, it is not expected for the coming generations of satellite terminals.

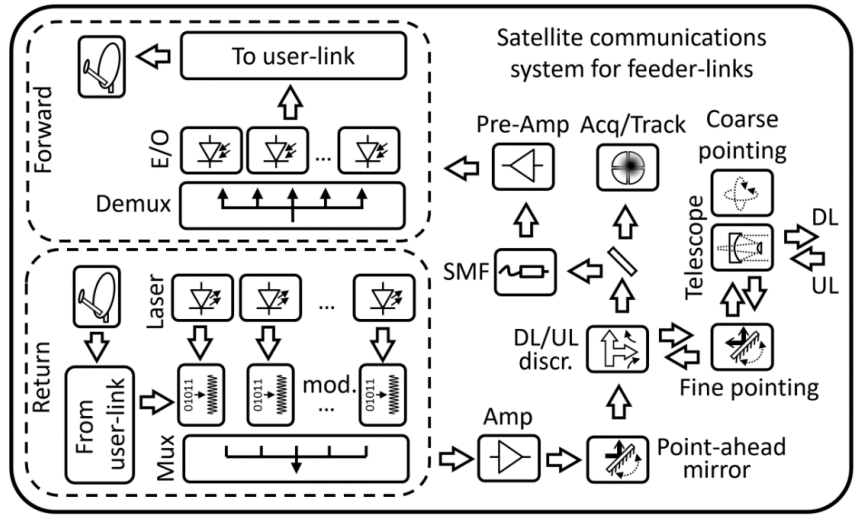

Fig. 8.33. Main components of the space terminal of a very-high throughput communications satellite based on optical feeder link.

Fig. 8.33 shows a block diagram with the main components of the space terminal of a very-high throughput satellite terminal based on optical feeder link. The telescope with the coarse and fine pointing units carry out the single-mode fiber coupling. The received signal is pre-amplified, demultiplexed, converted to the electrical domain, processed if required, and sent back after frequency conversion to the user link (here assumed to be in the Ka band). In the return channel, data from the user link is modulated on the multiple lasers, each one at a different center wavelength, multiplexed, amplified and after correction of the point-ahead angle, coupled into the telescope system and sent back to the ground station through the feeder link.

The forward link is the most challenging because of the following aspects:

- The atmospheric channel in the feeder uplink introduces signal fluctuations due to the combined effect of scintillation and beam wander, which produces a high dynamic range in the received power. Methods to minimize the signal fluctuations are implemented in the ground segment.
- Low-noise and high-sensitivity pre-amplifiers are a key technology for robust communications. The amplifiers must deal with the large dynamic range.
- Frequency up-conversion for the Ka-band user signals is required and represents an important part of the communications payload. Direct conversion from baseband is not applied to avoid non linearities. An intermediate frequency, like for example in C-band, can be used to simplify the onboard up conversion. However, this approach limits the capacity of the optical channel, becoming very inefficient. Another alternative is to perform the frequency conversion using optical technologies, based on coherent heterodyne reception [79]. This approach requires further optical components onboard the satellite, such as lasers, modulators and receivers, but it is quite straightforward to be integrated in an analogue transparent option since the signal can be directly converted in Ka band without further steps in the optical domain, simplifying the onboard system in a great deal.

- Onboard processing is the key for future digital satellite systems, especially when dealing with Terabit/s throughputs. Heat dissipation, power consumption and mass are design drivers of future space-qualified high-speed processors. Currently, this technology is still not available for such high-data-rate applications, but it may evolve in coming years.

For a relay scenario, the approach is similar, having only one optical channel and one RF channel and being the return channel the high-speed carrier of the data. In this case, the users are LEO satellites or HAPs (High-Altitude Platform), and they would make use of an inter-satellite link to transmit the data to the relay GEO satellite and the feeder link may be implemented in Ka band, as mentioned in 8.4.2. The main technical challenges are similar in this case, especially when increasing the data rate. The communications payload is however smaller than a system supporting feeder links and since there is an operational system already in space, new generations of such systems seem closer in time.

8.4.4 Ground segment

The main element in the ground segment is the telescope, whose main developments come from astronomy, where the telescopes diameter reached the 10 meters of diameter like the Grantecan in the Canary Islands, the Keck 1 and 2 in Hawaii or the Hobby-Eberly Telescope in Texas. The new generation of telescopes will reach the thirty meter class like ESO's Extremely-Large Telescope in Chile or the Thirty-Meter Telescope (TMT), planned to be built starting in 2020. For optical communications, small-class telescopes have been used. Experiments to GEO satellites have been performed with the 1.016-meter telescope of the ESA Optical Ground Station (ESA-OGS) in Tenerife, Canary Islands (see Fig. 8.34, left), and the 27-cm aperture of the Transportable Adaptive Optics Ground Station (TAOGS) (see Fig. 8.35).

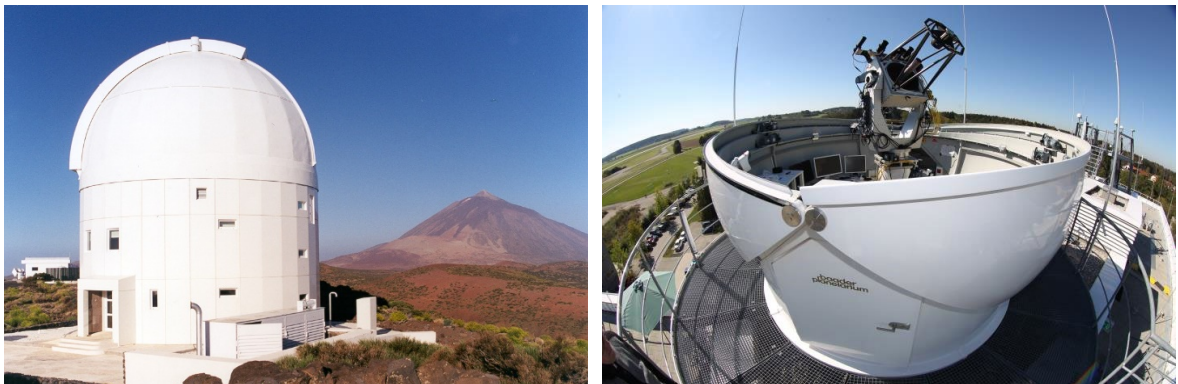

Fig. 8.34. ESA optical ground station in Tenerife, Canary Islands (credit ESA) (left). DLR optical ground station in Oberpfaffenhofen (right).

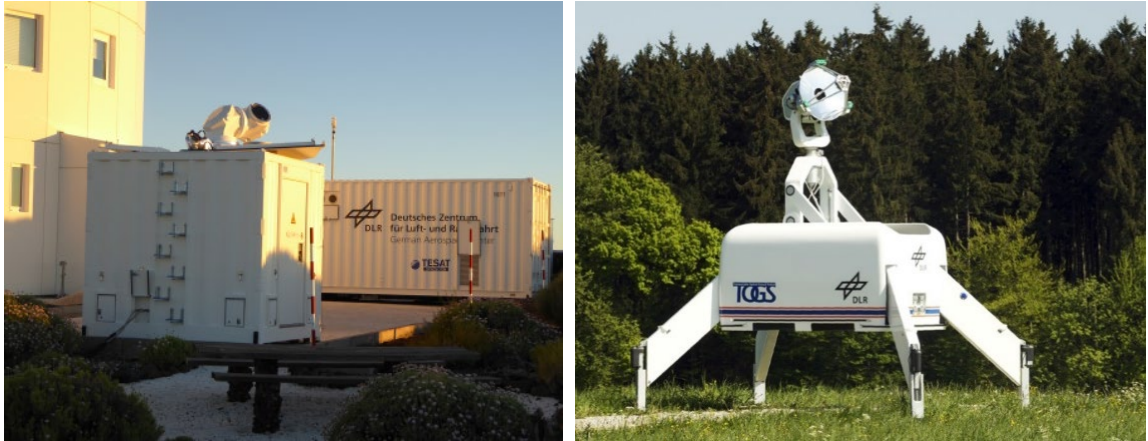

Fig. 8.35. DLR Transportable Adaptive Optics Ground Station (left). DLR Transportable Optical Ground Station (right).

The ESA-OGS in Tenerife (see Fig. 8.34, left) is located in the *Observatorio del Teide* in Tenerife, Canary Islands (Spain). The telescope is mounted on an equatorial mount and it has a Coudé path that allows redirecting the light from the telescope to an optical table. The telescope was built in 2001 to support ground-to-GEO satellite links with the optical terminal onboard the Artemis satellite within the SILEX project [55].

The Transportable Adaptive Optics Ground Station (TAOGS) (see Fig. 8.35, left) was developed by DLR and Tesat Spacecom to support ground-to-satellite optical links with the optical terminals onboard Alphasat and the EDRS satellites. This ground station is equipped with an adaptive-optics system that performs single-mode fiber coupling for data reception. All the equipment is installed in a container, which allows an easy transportation.

The DLR optical ground station in Oberpfaffenhofen (DLR-OP-OGS) (see Fig. 8.34, right) has a telescope on an azimuth-elevation mount with a 40-cm Ritchey-Chrétien telescope. The station will include an 80-cm telescope with Coudé path, which will be operational by 2020. The TOGS station (see Fig. 8.35, right) allows having a compact system, folding the 60-cm telescope into a carbon-fiber box. All these stations are typical examples of ground segment infrastructure for GEO satellite communications.

For the downlink, the large telescope diameter has a beneficial impact in the link budget, as the receiver gain increases with the receiver diameter, as shown in equation (8.17). This means that a larger receiver telescope may allow higher data rates in the downlink. However, an adaptive-optics system is typically required because the atmospheric turbulence limits the optical-signal coupling, as explained in more detail in the next subsection.

For the uplink, the transmitter size is limited also by the turbulence, and especially by the pointing accuracy, which is limited by the beam wander, as described by equation (8.35). When transmitting and receiving through different apertures, like in the DLR-OP-OGS, TAOGS and TOGS, the beam wander cannot be minimized, and therefore the divergence needs to cope with all the beam wander movements to make sure that the uplink reaches the satellite most of time, as shown in equation (8.40). The root-mean-squared value of the beam wander can be in the order of tens of microradians.

$$\sqrt{\sigma^2_{pointing}} = 0.73 \left(\frac{\sqrt{2}\lambda}{D_T}\right)\left(\frac{D_T}{\sqrt{2}r_0}\right)^{5/6} \qquad (8.40)$$

Transmitting and receiving from the same aperture allows sampling the tip-tilt in the right location, allowing a reduction of the beam wander and therefore a reduction of the transmitted beam divergence (involving an increase of the transmitting aperture) and a better power efficiency [80]. The main drawback of transmitting and receiving from the same aperture is the isolation between both directions. Circular polarization is typically used in satellite lasercom because it makes the system insensitive to rotations of the optical terminal when pointing or during the satellite movement. Therefore, left-hand and right-hand circular polarization are frequently used for downlink and uplink separation, however typical isolation of the polarization splitters/beam combiners falls around 20 to 30 dB. In order to increase the extinction ratio, a typical approach is to combine polarization with wavelength discrimination.

The main components of the ground segment for a very-high throughput communications system are shown in Fig. 8.36. The telescope transmits and receives the data to/from the satellite, the coarse-pointing system points towards the satellite and keeps the pointing error small enough to allow the fine-pointing assembly the compensation of the remaining angle-of-arrival fluctuations of the signal. The wavefront sensor (WFS) and the deformable mirror (DM) are part of the adaptive-optics system which compensates for the phase distortions of the atmospheric turbulence. Both adaptive-optics system and pointing system are applied for both link directions, in downlink for fiber coupling and in uplink for pre-compensation of the beam wander and phase distortions. In downlink, the light is coupled into a single-mode fiber, pre-amplified, demultiplexed, converted to the electrical domain for FEC and data processing before sending the data to the network. For the uplink, the data coming from the network is converted to the optical feeder-link format, modulated at each laser carrier, multiplexed, amplified and after compensating the point-ahead angle, coupled into the telescope system to transmit the signal towards the satellite.

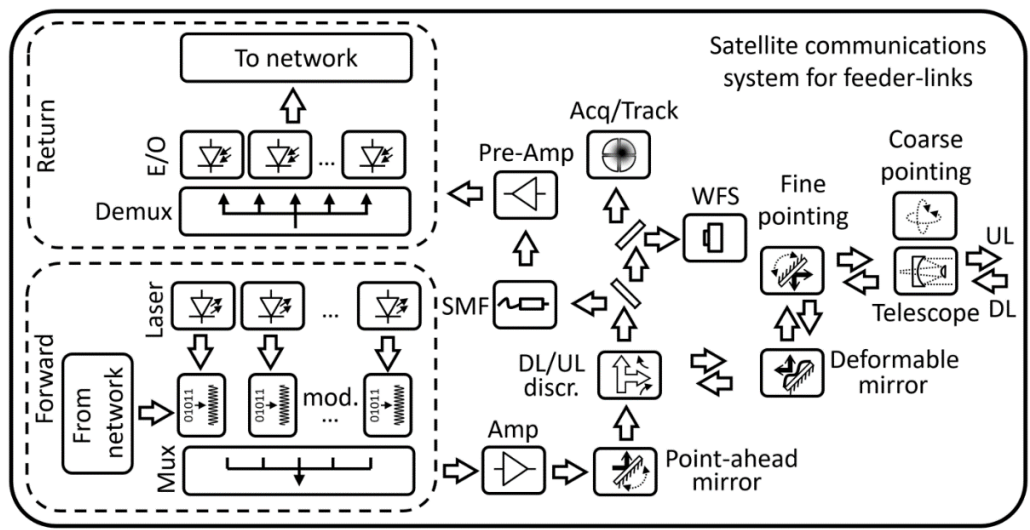

Fig. 8.36. Main components of the ground station for a very-high throughput communications satellite based on optical feeder links.

Some systems are required to make the ground station suitable for reliable and stable ground-to-GEO satellite communications. These systems are discussed in the following subsections:

- Adaptive optics for atmospheric-turbulence compensation
- Point-ahead angle and references for uplink pre-correction
- Aperture diversity for turbulence mitigation

Adaptive optics for atmospheric-turbulence compensation

The collected light by the telescope usually needs to be coupled into a single-mode optical fiber, to make use of all components developed for fiber communications like low-noise amplifiers or demultiplexers. At the ground station, the light wavefront is distorted due to the turbulence, limiting the performance of the fiber coupling. Phase distortions create intensity speckles at the focal plane of the telescope, which change size and position randomly. As a result, the coupling will usually add fading to the communications link. The coupling efficiency depends on the correlation between the received field and the coupling mode of the fiber. By correcting the phase distortions of the received field, the correlation with the coupling mode of the fiber increases, increasing stabilizing the amount of coupled light. The ratio between the aperture's diameter and the Fried parameter $D/r_0$ defines the "amount of phase distortions collected by the telescope". For telescopes smaller than the Fried parameter, only tip/tilt correction is required, since no phase aberrations are on the receiver aperture collecting area. But this implies low telescope gains, which has an impact in high-data rate links and GEO link distances, making such approach impracticable. By increasing the telescope diameter, adaptive optics is required and its complexity grows with increasing the telescope diameter.

The same requirements apply for coherent reception, where the incoming light is mixed with a local oscillator. To keep a high heterodyne efficiency, the received wavefront needs to match the local oscillator in order to demodulate the signal. Therefore, an adaptive-optics system is required in order to correct the phase distortions and keep a stable coupling or coherent reception. This kind of systems is widely used in astronomical telescopes to boost imaging performance towards the diffraction limit. The conditions for lasercom links are however different. While astronomy usually works around the zenith from astronomical sites (high locations), in communications lower elevation angles are targeted (current EDRS satellites are at around 30 degrees elevation from middle Europe), and ground stations are located also at lower altitudes. Only corrections of small fields of view are necessary in communications because the tracking keeps the counter-partner aligned, but the turbulence requirements are more demanding than in astronomy.

The main elements that constitute an adaptive-optics system are the tip-tilt mirror, the wavefront sensor (WFS) and the deformable mirror (DM). They are shown in Fig. 8.37. As explained before, the tip-tilt mirror compensates the angle-of-arrival fluctuations due to the atmosphere and any tracking errors, which cause spot shifts in the focal plane. The WFS estimates the phase of the received wavefront and a control computer computes the signals to drive the DM. The DM-surface shape is formed by a set of actuators and it is adapted to the received beam in order to conjugate (compensate) its phase distortion. For the concerned scenarios, the atmosphere can reach frequencies in the order of kHz, which causes very strong requirements on the adaptive optics system closed-loop speed.

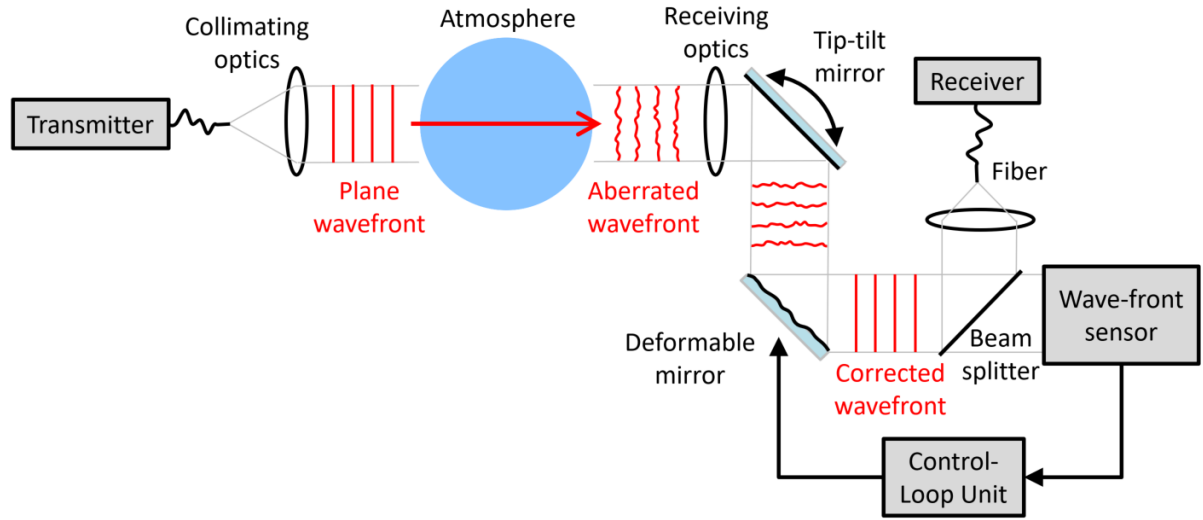

Fig. 8.37. Block diagram of an ideal adaptive-optics system.

An adaptive-optics system cannot perfectly compensate for the phase distortions, due to for example the limited number of actuators of the DM or the limited bandwidth of the control loop. As an example, from these two aspects, one can derive requirements for the control loop and for the required number of actuators. The number of actuators depends directly from the square of the ratio between the beam diameter and the actuator space, which has an impact on the residual phase error due to the limited resolution by fitting the received phase.

Recalling the definition of the Greenwood frequency of equation (8.31), which defines the characteristic frequency of the atmosphere, further requirements for the adaptive-optics closed loop can be defined. Similarly to equation (8.33), the residual phase-error due to the limited bandwidth of the closed loop is defined by equation (8.41) [81].

$$\sigma^2_{\theta,loop} = \left(\frac{f_G}{f_{AO,3dB}}\right)^{5/3} \qquad (8.41)$$

Another important element of the adaptive-optics system is the WFS, which estimates the phase of the received light. Typically, Shack-Hartmann sensors are used as WFS due to their relatively-simple hardware configuration. This kind of sensors is mainly used in astronomy, usually working around zenith, where the turbulence strength is lower [82]. This puts some limitations for communications were turbulence can become moderate to strong. Shack-Hartmann sensors are based on the measurement of the phase gradient by means of an array of lenses placed at the pupil. The focus of each lens is imaged by a camera and, by means of centroid or a correlation-based algorithm, the phase is reconstructed. The main limitation of such approach appears when turbulence increases. In this case, some of the sub apertures do not receive enough light due to scintillation, introducing errors in the reconstruction. Furthermore, the appearance of a rotational component of the phase under strong turbulence has been studied, concluding that it cannot be observed by sensors based on the measurement of the phase gradient, like Shack-Hartmann or curvature sensors [83]. As an alternative, interference based WFS, iterative methods or the combination of two sensors are options to be considered to achieve a more resistant wavefront estimation under strong turbulence [84] [85].

Point-ahead angle and references for uplink pre-correction

In Fig. 8.38, there is a representation of a bidirectional link between a satellite and a ground station. The ground station receives the downlink and tracks the incoming light. Due to atmospheric turbulence, a tip-tilt mirror is required for compensating the angle-of-arrival fluctuations. In the meantime, the satellite moves, and the ground station needs to point-ahead by a certain angle, as explained in sections 8.1.2 and 8.3.3. This point-ahead angle for a GEO satellite is around 18 µrad. The pointing direction of the uplink will fluctuate due to the atmospheric turbulence, the so-called beam wander, as explained in section 8.2.2 and described by equation (8.28). Using the measurement of the angle-of-arrival fluctuations, the uplink could ideally be compensated against beam-wander by applying the same fluctuations as a pre-compensation. Unfortunately, due to the point-ahead angle, both uplink and downlink travel through different atmospheric paths. The atmospheric effects are however correlated within a certain cone angle, called isoplanatic angle, as represented by the green triangle in Fig. 8.38. The isoplanatic angle takes into account the correlation of all phase distortions. There is a similar definition for only the tip/tilt effects, called isokinetic angle. How large both angles are depends on the turbulence conditions.

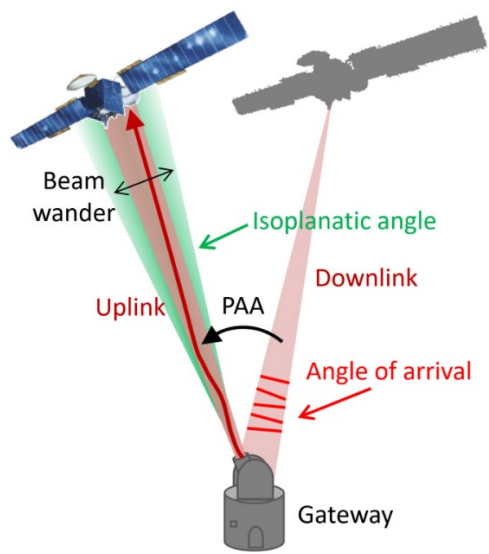

Fig. 8.38. Isoplanatic angle and point-ahead angle in satellite bidirectional links.

As a rule of thumb, isokinetic angle is about 1.5 to 2 times larger than the isoplanatic angle. Meaning that, higher aberration orders of the phase-distortions have smaller coherence angles and accordingly, also shorter coherence times. This is however not an ON/OFF process. Being outside the coherence cone of the atmosphere means an increase of the decorrelation between both paths, and therefore a higher residual beam wander, which leads to higher intensity fluctuation at the satellite. The larger the separation between uplink and downlink, the larger the decorrelation, until both paths are completely independent. The most straightforward way to mitigate beam wander is to increase the divergence, as pointed out by equation (8.35). The main drawback is the decrease on the mean power at the satellite, due to a larger divergence beam. At the end, this is a link budget optimization between the beam divergence, the probability of outages and the received power, as applied in [77].

Another alternative is to have a reference in the direction of the uplink, as used in astronomy with laser guide stars [86]. Laser guide stars based on Rayleigh scattering have been studied since beginnings of the nineties in astronomy [87] [88]. The main drawback of Rayleigh scattering is that this effect happens mainly in the first kilometers of the propagation, limiting the adaptive optics performance due to the short distance, at which the reference signal is located [89]. The most promising technology is the laser guide stars based on the sodium atom excitation in the mesosphere, in the sodium layer of the atmosphere, about 90 km height [90]. By transmitting a laser at 589 nm, the sodium atoms in this layer get excited and produce an "artificial star" that can be used as a reference for adaptive optics. In astronomy, this laser is transmitted within the isoplanatic angle of the observation direction, in order to apply adaptive optics on the observed object. In communications, this technique could be applied in the direction of the uplink, in order to apply pre-correction adaptive optics [77].

As described in Fig. 8.37, adaptive optics corrects for the phase distortions in the downlink in order to enable single-mode fiber coupling. The same system can be used for transmitting a laser in the other direction, pre-distorting the phase. The goal is to decrease the intensity fluctuation produced by the atmospheric phase distortions. Since the isoplanatic angle may be smaller than the point-ahead angle, a laser guide star could provide a reference in the uplink direction that could be used for correcting the phase. The main point to be solved remains in the tip-tilt correction. By observing the laser guide star from the transmitter direction, no tip-tilt can be measured because the light travels up and down through the same path. In order to use the laser guide star for tip-tilt correction an off-axis observation is required, as proposed in [91]. Currently, there are experiments targeting a demonstration of real-time compensation of the beam-wander by means of laser guide stars [92].

Aperture diversity for turbulence mitigation

The integration of an adaptive-optics system at the ground station is not the only approach for minimizing the turbulence impact. The use of transmitter diversity and receiver diversity is also a possibility and it has been applied, at both pupil and focal plane.

At the pupil plane, spatial diversity exploits the decorrelation between different optical paths through the atmosphere in transmission or reception. As rule of thumb, for transmitted beams, the atmospheric paths are assumed uncorrelated when they are half a meter a part. In this case, the transmitted beams overlap after some km of transmission, so that the receiver on the satellite receives only the combination of all the beams, hence fluctuations average out. Statistically, scintillation decreases linearly with the number of transmitted beams, when each path is assumed independent of each other.

This approach works properly when no data is modulated on the laser and the signals come from independent laser sources, for example, for uplink beacons. If only one single-mode source is used, for example, by splitting the signal between transmitters, the combination of the signals at the satellite can lead to interference patterns, because the path difference between them may lie within the coherence length of the laser. Applying a delay between transmitters may be an option, for example by transmitting the light through some kilometers of fiber for example.

In the SILEX ground station, 4× transmitter diversity was implemented (four transmitted uplink lasers separated around half a meter) using one single multimode laser, but the results were not the expected [93]. In the case of multimode lasers, the visibility of the interference pattern is a periodical function of the path difference between laser signals; therefore, delay between transmitters needs to be accurately selected [80].

In case data is modulated on the laser carrier, the bandwidth of the signal increases. An overlap of even only a part of the bandwidth will lead to a strong interference. In this case, the signals may be separated in polarization or wavelength for example. This can be done because the chromatic dispersion around the communications wavelength is very low and the polarization remains unchanged by transmitting through the atmosphere. But both approaches are not really advantageous. Usually, polarization (in combination with wavelength) is used for the separation between link directions. And wavelength separation (diversity) requires the data spectrum times the number of transmitted beams, which is unfeasible for very-high throughputs. More elaborated approaches like generating single-side bands may be an alternative to place two transmitters in the same bandwidth [94], but its application for coherent communications schemes need to be still investigated since most of these approaches requires non-coherent detection. Alamouti transmitter diversity scheme would be the best option, being compatible with coherent modulation schemes, but the receiver complexity at the satellite increases.

For the reception, diversity can be applied by an array of telescopes. This approach has been mainly thought and test for deep-space scenarios, especially to avoid very large monolithic mirrors. For example, in the lunar link within the LADEE mission, where the LLGT telescope was developed. In this case, photon-counting technology was used, and the four received signals were combined in the electrical domain.

In case the telescope apertures are small enough, adaptive optics could be avoided and only tilt-correction is needed. However, signals need to be combined either electrically or optically, which increase the hardware complexity at the receiver side, by aligning the phases of the received signals optically or by increasing the number of receivers. In [95], receiver diversity is shown by combining four coherent receivers electrically. The combined signal shows the reduction of the atmospheric fluctuations. This assumes however that the aperture diameter of each telescope is big enough to be within the signal sensitivity, which limits the minimum diameter of the telescope, which is again a trade-off with the expected turbulence, since we are assuming that no adaptive optics is needed.

The same approach of diversity could be followed at the focal plane, by, for example, setting a fiber bundle. In presence of turbulence, the intensity distributes randomly at the focal plane: several speckles change position and intensity. A fiber bundle can collect more power and combine the signals optically or electrically, following the same approach than before. Such approaches have been developed to increase the FOV of lasercom systems [96]. The same idea can be though by using a multimode fiber. In [97], a multimode fiber to multiple single-mode fiber was developed and demonstrated. In this case, the phase distortions may excite several modes in the fiber and each one is coupled in different single-mode fibers. Another idea is to bring back the optical power distributed among the different modes to the main mode, since most of components developed for fiber communications, in particular related to DWDM, are based on single-mode fiber.

8.5 Future optical satellite networks

A satellite network can be defined by a set of satellites at different orbits, combining GEO, MEO and LEO, for exploiting the visibility of the higher orbits and the short delays of the lower orbits. Although in the past there were a lot of studies on defining satellite network architectures and their traffic management, current systems are focused in only one orbit: mainly LEO and GEO. Announced systems will complement the satellite services with further GEO satellites and constellations and LEO and MEO, focusing mainly in providing internet.

In the following sections, there is a short description of the current satellite systems, the main applications and finally some considerations on the network architecture.

8.5.1 Current and upcoming satellite systems

Looking at the current systems at the different orbits, in GEO the satellite communications market is mainly focused on video broadcasting and internet access and based on Ka-band and Ku-band technology. Some examples of HTS satellites in service are Ka-Sat with 70 Gbit/s and Eutelsat 172B, based on Ku-band technology and operating in Asia Pacific; ViaSat1 and ViaSat2 with 140 Gbit/s and 350 Gbit/s; Inmarsat Global Xpress, which are four GEO satellites offering global coverage; Intelsat (Epic Series); GSAT-19 covering India; etc.

MEO orbit is used primarily from Navigation systems like GPS, Galileo, Beidou (with also GEO satellites), Glonass. This orbit is also used by the O3b constellation for broadband communication. 16 satellites provide 16 Gbit/s per satellite with 700-km beam footprints. A MEO constellation is proposed by Laserlight Communications [98], who is planning to deploy an all-optical constellation with 12 MEO satellites in collaboration with Optus.

LEO constellations are traditionally used for Mobile Satellite Service, both with global access as for the fully inter-satellite and inter-orbit meshed IRIDIUM network, or regionally with bent pipe transponders as for the Globalstar and Orbcomm networks.

Currently, LEO-satellites constellations offer relatively-low data throughput. Examples are:

- Iridium next generation is constituted by 66 satellites providing L-Band communication to mobile users (128 kbit/s), up to 1.5 Mbit/s to Iridium Pilot marine terminals, and high-speed Ka-band service up to 8 Mbit/s to fixed/transportable terminals.
- Globalstar second-generation constellation will consist of 24 satellites and it offers mobile-voice communications and low data rate transfer. This system is used in monitoring areas like oil and gas, government, mining, forestry, commercial fishing, military applications or transportation.
- Orbcomm is 100% dedicated to M2M communications. The constellation is constituted by 50 satellites.

New developments are currently taking place on the LEO constellations:

- LeoSAT is a satellite constellation with around 100 satellites planned, which aims at deploying broadband services with user access rates from 50 Mbit/s up to 1.2 Gbit/s, with optical inter-satellite links and user/gateway links in Ka band, and on-board processing on each satellite.
- Starlink is a satellite constellation developed by SpaceX with the support of Samsung, with around 12,000 satellites planned, with optical inter-satellite links and ground receivers with phased-array antennas.
- Oneweb plans to have about 900 satellites, providing Internet access world-wide. The system is foreseen to be operational in 2019. The ground receivers will spread the Internet connection using 3G, 4G, 5G or WIFI.

8.5.2 Applications for future optical satellite networks

A variety of applications would profit of a global satellite network, where optical frequencies can play a key role, increasing the data throughput and allowing freeing currently allocated RF frequencies. The main applications are divided into terrestrial, maritime and aeronautical.

Broadband Internet access for private users and for industry is the main terrestrial application, which is currently strongly pushed in Europe. The European Commission set the objective of 30 Mbit/s per EU home in 2020 in the Digital Agenda. In a study of the broadband coverage in the EU [99], the coverage outside the big cities is 30% in most of EU countries. Fiber optics to serve the needs of areas with sparse population is an expensive infrastructure. Satellite communications may be a complementary solution to the current existing infrastructure. In this case the requirement is 30 Mbit/s per user or more. For such scenario, optical feeder-links seem a good opportunity, because of the massive non-regulated spectrum.

5G targets very-low latency and high capacity using high-frequency bands but the coverage at these frequencies is very limited and they are foreseen to be as close as possible to the end user to minimize the latency. The coverage decreases with the latency and inversely with the capacity. The bands are assigned to the new 5G applications and the legacy to support 2G, 3G and 4G [100]. Therefore, current deployment of the 5G network would benefit from a satellite system that allows setting new stations outside the optical-network coverage. The 5G network targets to revolutionize the industry with the so-called Industry 4.0. monitoring and telecontrol of production. A particular interest of optical communications in this field is the use of the optical frequencies itself which avoid potential frequency overlaps.

Tactile Internet would also benefit of broadband satellite communications. This concept includes applications like Online Gaming, Virtual reality, Robotics and Telepresence, with latency constraints down to 1 ms and capacity constraints up to 1 Gbit/s, or factory automation, heath care, with latency constraints down to 10 ms and capacity constraints up to 100 Mbit/s [101] [102], and these are only some examples.

Current latency measurements of the terrestrial network latency, provided by [103], show values of around 11 ms within Europe, below 75 ms for transatlantic communications or around 100 ms for transpacific, as shown in Table 1 2. These values can be compared to the round-trip latencies to the satellites in Table 8.6, assuming ground-to-satellite links at 10-degrees elevation. As expected, LEO satellites can offer lower latency due to their proximity to the Earth, however the very-low latency constraints of the Tactile Internet cannot be respected by any satellite connection, but also not by current terrestrial infrastructure.

Table 8.6. Latency in ms for terrestrial Internet connections and round-trip satellite connections.

| Terrestrial Fiber Network [103] | |
|---|---|
| Trans-Atlantic | 73.2 |
| Europe | 11.2 |
| North America | 36.6 |
| Intra-Japan | 9.3 |
| Trans-Pacific | 102.1 |
| Asia Pacific | 97.9 |
| Latin America | 131.5 |
| EMEA to Asia Pacific | 130.6 |
| Satellite round-trip | |

| LEO | 18.4 |
| MEO | 93.5 |
| GEO | 272.1 |

A Mobile Edge Cloud close to the end users should provide the cloud-based platform for all applications with very-low latency. In this architecture the latency constraints are set in the connection between the users and the Mobile Edge Cloud, by optimizing the communications physical layer. "Caching and in a more general category, information centric networking, can be assumed as one of the promising candidate technologies to design a paradigm in a shift for latency reduction in next generation communication systems" [102]. The satellite communications may play a role in the connection between the core network and the mobile edge cloud.

For maritime applications, tracking and monitoring of vessels, people and goods and sharing information between vessels is becoming essential and reliable worldwide communication for vessels is required. Especially polar areas, and particularly in the Artic above the 76°N, are not well covered. Future maritime transport of goods would require better communication services to allow receiving updated information on the route (e.g. ice, currents, weather) and keep tracking of the vessels and goods. In the future, autonomous transport or remote control of the vessel would allow covering the needs on the growing goods transport.

The main use cases are [104]:

- e-Navigation: This a project of the International Maritime Organization (IMO) towards a future digital concept for the maritime sector. This project includes on-route information updates (destinations, waypoint, and route optimization, maps update, weather information, ice information, or currents information), onboard low-cost communication services for crew entertainment and the Automated Information System (AIS) satellite-based data exchange system using the commonly-carried VHF equipment for complementing the terrestrial network access. In this respect, the VHF Data Exchange System (VDES) will extend the coverage also in artic regions. The expected traffic is in the order of 400 kbyte/hour/ship. This system is depending on the approval of the WRC19.
- Arctic, and Polar Regions in general, are not well covered by current communication systems. Communication using GEO satellites is theoretically possible up to 81°N, but typically only up to 76°N. Inmarsat satellites serve the Arctic up to 76°N except for an area around 120°E of the Laptev Sea Russia and around 120°W of the Beaufort Sea Canada. The polar orbiting satellites of Iridium and Cospas-Sarsat serve the whole of the Arctic. IMO has procedures in place to possibly recognize satellite systems in addition to the systems provided by Inmarsat and Cospas-Sarsat.
- Autonomous ships: unmanned merchant ships on intercontinental voyages with advanced sensor systems to detect and avoid obstacles, and allow advanced onboard control, positioning and navigation system to determine and control exact location, speed and course as well as route. In this case, a latency of less than one second is required so that the ship can be remotely controlled in real time. A bandwidth of up to 4 Mbit/s is required to send radar and video pictures.
- The Global Maritime Distress and Safety System (GMDSS) is a radio system whose techniques and frequencies are defined by the ITU and for which mandatory equipment carriage

> requirements have been adopted by the IMO for commercial vessels. A satellite network mal provide alternatives to High-Frequency (HF) (3 to 30 MHz) communications, where the traditional means of long-distance communication for ships are still used as a means of backup to GEO-satellite services.

In aeronautical communications, a modernisation of the current air traffic management (ATM) system is expected. The ATM system manages all aircraft in controlled airspace. The system uses analogue double-sideband amplitude modulation (DSB-AM) deployed in the VHF band between 118 and 137 MHz to communicate with the pilots. This modulation is very spectrum-inefficient and voice communications cannot cope with the needs of increasing air traffic. As predicted by the EUROCONTROL Statistics and Forecast Service [105], air traffic will increase by 50% by 2035, which will bring the current ATM system to its limits, especially in the most dense flight regions like Europe and United States. In the 1990s, the International Civil Aviation Organization (ICAO) standardized the VHF data link (VDL) standards, a digital system with 25-kHz bandwidth, allowing data communications. However, the link capacity is well below the requirements of aeronautical communications.

ICAO recommended the use of the L band between 960 and 1164 MHz, but ensuring the coexistence of already existing systems. In this frame, the L-band Digital Aeronautical Communications System (LDACS) was developed based on orthogonal frequency-division multiplexing (OFDM) together with adaptive coding and modulation, exploiting the 500-kHz bandwidth available in L band. The communication ranges between 561 kbit/s, using strong coding and robust modulation, and 2.6 Mbit/s, using higher modulation orders and weak coding [106]. LDACS standardization is currently under way in ICAO and it is planned to start in 2018.

A satellite-based communication solution for the European ATM System is driven by ESA in Iris, the ARTES programme Satellite Communication for Air Traffic Management, in partnership with Inmarsat. High-capacity digital data links via satellite could allow transferring information of latitude, longitude, altitude and time to monitor and adjust the aircraft route efficiently, when necessary, like due to change on weather conditions. The target is 2028, when Iris will support the service around the globe.

In summary, aeronautical applications can profit from direct broadband access to a global network supporting onboard WLAN (passenger aircraft) or real-time transmission of reconnaissance data (governmental and disaster relief missions). Similarly, maritime applications will profit from bandwidth enhancements with regards to existing narrowband and medium-bandwidth services such as INMARSAT's BGAN services. Land applications are expected to include broadband fixed and nomadic access as well as intermediate and low-bandwidth mobile access. Industry 4.0 and IoT applications are expected to be supported as well, e.g. connecting dispersed sensors, control and command of remote installations like satellite ground terminals, fleet management etc.

8.5.3 Network architecture

Nowadays, optical satellite communication has been developed for point-to-point data transmission, either through direct LEO downlinks or through data relay over GEO. But space lasercom can potentially have a huge impact on enlarging the ground network infrastructure to space, achieving global coverage of areas lacking ground infrastructure with broadband access.

From the radiofrequency systems point of view, satellite networks have been investigated for a long time and schemes have been proposed combining different constellations in LEO and MEO [107], or combining GEO, MEO and LEO [108]. Routing in satellite networks have been studied defining several layers. For example, in [108], MEO satellites manage the traffic in the LEO constellation within their footprint and GEO satellites monitor and manage the whole system, informing the gateways about traffic congestions or inter-satellite links failures. The design of the switching and monitoring approach among layers to guarantee a desired QoS (Quality of Service) depends on the number of layers (constellations and satellites per orbit), the number of inter-satellite links and the available throughput. The design of such a system may need to consider a combination of radiofrequency and optical links to cope with all the possible applications.

LEO-satellites constellations have been considered for a long time due to the advantage of short delays, favourable power budgets, and closer distances compared to MEO and GEO. For satellite-to-ground optical communications, this last aspect is at first sight no more a clear advantage because of the limited pointing accuracy due to the larger point-ahead angles at LEO and MEO, which prevents the beam-wander compensation, as discussed in Fig. 8.38. As a rule of thumb, one may need a beam divergence 5-10 times larger for LEO communications than for GEO. The main drawback is therefore a decrease of the received power, leading to similar transmitted power requirements for LEO and GEO, with current technologies. The advantage of having LEO closer is lost because of the point-ahead angle. Furthermore, LEO-satellite payloads will be smaller due to the larger number of satellites, thus the constellation may rely on a low-cost deployment. That means that telescopes onboard the satellite will tend to be smaller, having also penalties on the receiver gain. In case of satellite-to-ground radiofrequency links, multiple active gateways must be active to achieve high throughputs.

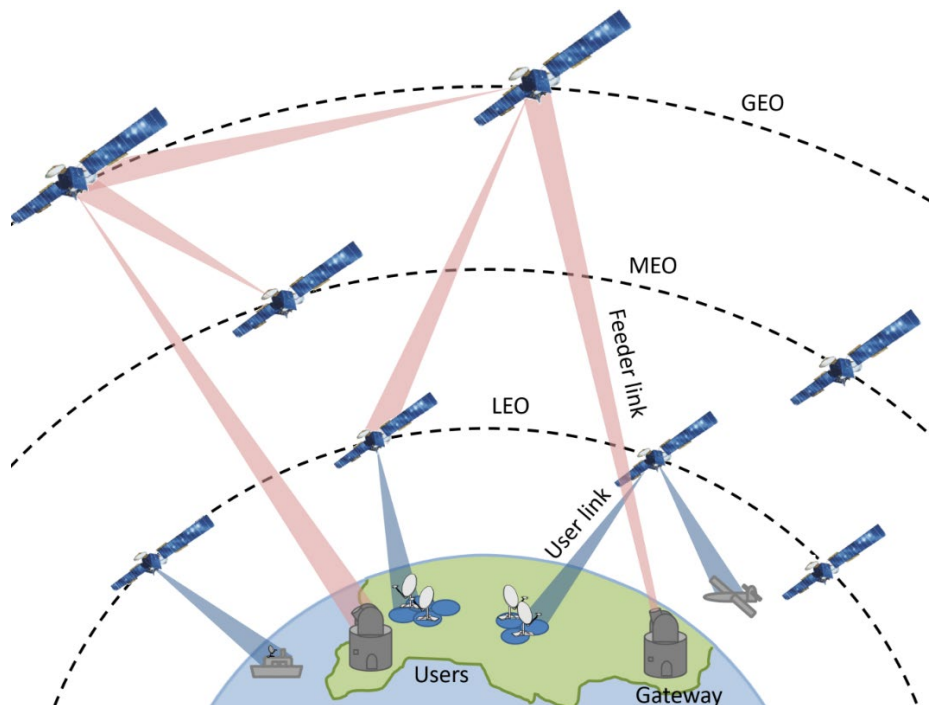

Fig. 8.39. Optical satellite-communications network.

GEO satellites are primary more expensive in the development and deployment, because of the larger distance to be covered in launch and the higher requirements due to radiation. The main advantage of GEO satellites is the large coverage that achieves one satellite and due to the properties of the atmosphere, the small divergences of the optical beams can be tightened narrower than in LEO or MEO. Most likely, an optical global network will be a combination of the three orbits (as represented in Fig. 8.39), optimizing the QoS and combining several applications, maybe even including navigation services in MEO or in LEO, as proposed by the Stanford University [109].

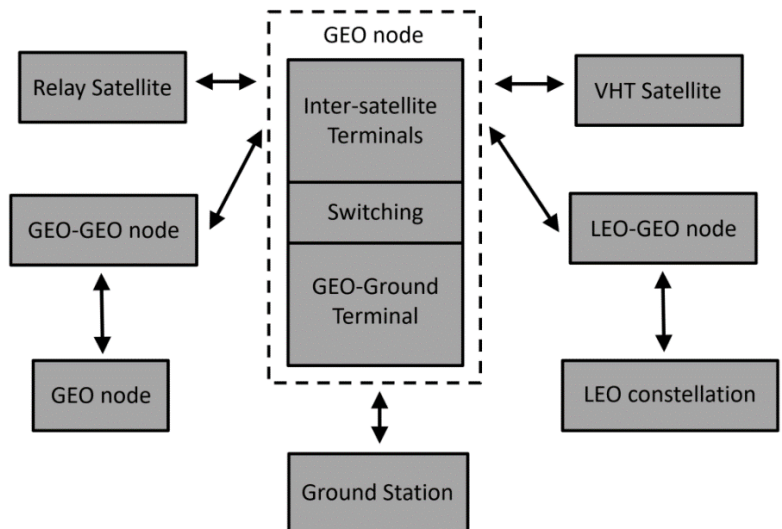

Fig. 8.40. Conceptual block diagram for an optical satellite network.

A system combining the variety of applications may need to combine a mesh-configuration with satellite constellations. Fig. 8.40 shows a concept for a satellite network. The development of ad-hoc satellite platforms, for backhauling and switching between ground and GEO for example would allow increasing the data throughput of the feeder links. In this case, the GEO platform could incorporate signal regeneration and optical switching towards other application-oriented satellites. This is especially interesting for onboard signal processing, for example for error-correcting algorithms protecting the data across the atmospheric turbulence. GEO satellites at shorter distances can be easily connected with optical links at limited power requirements and transporting high-data volumes. Application-oriented satellites, for example based on data-relay or VHT communications, could have simplified payloads, fostering their own applications. Other dedicated platforms may connect other GEO nodes at large distances, for example between Europa and Asia or to LEO constellations. Optical frequencies are the best candidate for such networks: they are more efficient in mass and power, they are more resistant to interferences and it solves the RF spectrum bottleneck. However, a combination of the optical and RF technologies is required to satisfy the requirements of such a vast variety of applications.

<i>Alberto Carrasco-Casado, Ramon Mata-Calvo "Free-space optical links for space communication networks"
https://doi.org/10.1007/978-3-030-16250-4_34</i>